\documentclass[fleqn,10pt]{wlscirep}

\usepackage[T1]{fontenc}
\usepackage[utf8]{inputenc}
\usepackage{currvita}
\usepackage{xcolor}
\usepackage{graphicx}%
\usepackage{multirow}%
\usepackage{amsmath,amssymb,amsfonts}%
\usepackage{amsthm}%
\usepackage{mathrsfs}%
\usepackage{textcomp}%
\usepackage{manyfoot}%
\usepackage{booktabs}%
\usepackage{algorithm}%
\usepackage{algorithmicx}%
\usepackage{algpseudocode}%
\usepackage{listings}%
\usepackage{makecell}
\usepackage{tabularx}
\usepackage{adjustbox}
\usepackage{array}
\usepackage[title]{appendix}%
\usepackage{hyperref}
\usepackage{caption}
\usepackage{threeparttable}

\renewcommand{\thefigure}{\thesection\arabic{figure}}
\usepackage{float}
\let\oldequation\equation
\let\oldendequation\endequation

\renewenvironment{equation}
{\linenomathNonumbers\oldequation}
{\oldendequation\endlinenomath}
\usepackage{lineno}

\title{Intensification of Oceanic Inverse Energy Cascade Under Global Warming}

\author[1]{Qianqian Geng}
\author[1,*]{Ru Chen}
\author[2]{Bo Qiu}
\author[3,4]{Zhao Jing}
\author[1]{Xin Su}
\author[5]{Gang Huang}
\author[6]{Qinyu Liu}
\author[1]{Yang Chen}
\affil[1]{Tianjin Key Laboratory for Marine Environmental Research and Service, School of Marine Science and Technology, Tianjin University, Tianjin, 300072, China}
\affil[2]{Department of Oceanography, University of Hawaii at Manoa, Honolulu, 96822, USA}
\affil[3]{Frontiers Science Center for Deep Ocean Multispheres and Earth System and Key Laboratory of Physical Oceanography, Ocean University of China, Qingdao, 266100, China}
\affil[4]{Laoshan Laboratory, Qingdao, 266100, China}
\affil[5]{State Key Laboratory of Numerical Modeling for Atmospheric Sciences and Geophysical Fluid Dynamics (LASG), Institute of Atmospheric Physics, Chinese Academy of Sciences, Beijing, 100029, China}
\affil[6]{Physical Oceanography Laboratory, College of Oceanic and Atmospheric Sciences, Ocean University of China, Qingdao, 266100, China}

\affil[*]{Corresponding author. Email: ruchen@tju.edu.cn}

\begin{abstract}
Kinetic energy (KE) cascade in the turbulent ocean is pivotal in connecting diverse scales of oceanic motions, redistributing energy, and influencing ocean circulation and climate variability. However, its response to global warming remains poorly understood. Using a 24-year satellite altimetry dataset, \textcolor{black}{we identify a pronounced intensification of inverse  geostrophic kinetic energy cascade at the sea surface across most ocean regions during 1994-2017, with cascade amplitude increasing by 1$\%$ to 2$\%$ per decade. This intensification occurs not only in energetic regions but also in expansive quiescent areas. Contributing factors to this intensification of geostrophic KE cascade include enhanced vertical shear of horizontal velocity, deepened mixed	layer, strengthened stratification, weakened eddy killing as well as changes in the KE budget. The dominant factors vary across regions. Our findings offer new insights into the ocean’s response to global warming and improve understanding of feedback mechanisms for ocean circulation changes.}

\end{abstract}
\begin{document}

\flushbottom
\maketitle

\pdfoutput=1
\thispagestyle{empty}

\textcolor{black}{The rapid increase in carbon emissions and the resulting global warming in recent decades have induced significant and persistent changes in the world's ocean \cite{bakun1990global,lee2023ipcc,hanna2008increased,collins2010impact,toggweiler2008ocean,mecking2023decrease}, particularly  in oceanic kinetic energy (KE) \cite{2020Deep,wang2024more,Josu2021Global,beech2022long,peng2022surface}. Initially \textcolor{black}{driven by} atmospheric and tidal forcing, oceanic KE sustains a wide range of motions spanning scales from  O($10^4$) km down to O(1) mm, which include large-scale overturning circulation, western boundary currents, mesoscale eddies and other smaller-scale motions \cite{2009Ocean,storer2022global,balwada2022direct,storer2023global}. These motions play a critical role in regulating the global climate system \textcolor{black}{through influencing the distribution of heat \cite{zhang2014oceanic,dong2014global,wu2012enhanced,rahmstorf2002ocean,oceanrole2018,hu2015pacific}, carbon storage \cite{ito2015sustained,southernocean2024,lovenduski2008toward,le2007saturation,gnanadesikan2015isopycnal}, etc.} As such, changes in KE could have profound impacts on \textcolor{black}{the global climate system.} Given the constant interactions among multi-scale motions, KE is intensely transferred across scales before dissipating at mixing scales, enabling the equilibration of global ocean circulation \cite{2009Ocean,qiu2022bi}. KE cascade, a fundamental process in turbulence \cite{kolmogorov1941local,2006Atmospheric,2018Cascades}, plays a vital role in mediating multi-scale energy transfers and contributes to the long timescales of climate variability \cite{serazin2018inverse,storer2023global}. Studying changes in KE cascade thus offers insights into changes in multi-scale oceanic motions, enhances our understanding of the ocean’s capacity for heat and carbon uptake\cite{rahmstorf2002ocean,oceanrole2018,wu2012enhanced,hu2015pacific,dong2014global, ito2015sustained,southernocean2024,lovenduski2008toward,le2007saturation,gnanadesikan2015isopycnal}, and advances efforts to tackle climate change challenges.}  

\textcolor{black}{Despite extensive research on the dynamics and characteristics of KE cascade  \cite{2005Direct,Aluie2018Mapping,tulloch2011scales,wang2015geographical,dong2020seasonality,li2021barotropic,garabato2022kinetic,steinberg2022seasonality,storer2023global}, \textcolor{black}{its evolution under global warming remains largely unexplored. Recent studies on the ocean’s response to global warming have primarily \textcolor{black}{focused} on individual scales of oceanic motion, such as the acceleration of large-scale ocean circulation\cite{2020Deep,peng2022surface} and the global evolution of mesoscale currents \cite{Josu2021Global,beech2022long}.} \textcolor{black}{Little is known about the changes in multiscale energy interactions and, consequently, energy cascades.} \textcolor{black}{Recent literature has shown that over the past several decades, stratification\cite{li2020increasing,sallee2021summertime}, mixed layer depth\cite{sallee2021summertime}, and vertical shear of ocean circulation\cite{wang2024more} have all significantly increased across the global ocean. \textcolor{black}{Considering energy cascade, mesoscale eddy properties and baroclinic instability processes are all sensitive to the background mean flow shear, stratification and layer depth \cite{arbic2000generation, 2006Atmospheric,  scott2007spectral,tulloch2011scales,venaille2011baroclinic,jia2011interannual,wu2023seasonal,feng2022seasonality}}, we hypothesize that under global warming, KE cascades have undergone substantial changes as well.}} 

 Here, we evaluate the trends in the global surface geostrophic KE cascade using satellite altimeter data spanning from 1994 to 2017. \textcolor{black}{As noted in a recent study \cite{qiu2014seasonal}, the strength and scale of the KE cascade can be significantly influenced by oceanic processes occurring at different spatial resolutions, such as mesoscale and submesoscale baroclinic instabilities. Given that the satellite altimetry primarily resolves mesoscale to larger-scale features, this study focuses on the KE cascade from mesoscale to large scales, approximately ranging from O(60) to O(1000) km.} The spectral KE flux, an effective metric to measure KE transfer\cite{2005Direct}, is diagnosed using the coarse-graining approach \cite{Aluie2018Mapping,storer2023global,garabato2022kinetic,schubert2020submesoscale}. Given the dominance of the inverse KE cascade in the ocean\cite{2005Direct,storer2023global,wang2015geographical}, our analysis focuses on trends in key inverse KE cascade characteristics, including cascade amplitude, injection scale, arrest scale and amplitude scale (Fig. \ref{fig1}b). We present robust evidence of a pronounced inverse cascade enhancement in both energetic ocean regions and extensive quiescent regions over the period 1994-2017. This enhancement is largely consistent with changes in mixed-layer depth, stratification, vertical shear of horizontal velocity and eddy killing, along with adjustments in the KE budget.

\section*{Global inverse KE cascade and its strength}
 We depict the strength of inverse KE cascade using the metric $Amp$, which is defined as \textcolor{black}{the amplitude of inverse KE cascade} (see Fig. \ref{fig1}b and `Metrics of KE cascade characteristics' in Methods). This metric reflects the ocean’s capacity to transfer KE upscale. Consistent with previous studies \cite{wang2015geographical,li2021barotropic}, robust inverse KE cascade typically occurs in eddy-rich and highly energetic regions, such as the western boundary currents and the Antarctic Circumpolar Current (Fig. \ref{fig1}a). In equatorial regions, areas with no inverse KE cascade are masked, likely due to the prevalence of 3D turbulence flow in these areas \cite{storer2023global}. Given that the energetic ocean regions have significant climate impacts\cite{hu2015pacific,cai2023southern,kelly2010western,kubota2014larger} and exhibit prominent inverse KE cascade (Fig. \ref{fig1}a), we next divide the global ocean into several high-energy regions and one quiescent region for regional cascade trend analysis (Fig \ref{fig2}). 
  
 \renewcommand\thefigure{\arabic{figure}}  
 \renewcommand{\figurename}{Fig.}
 
 \begin{figure}[ht]
 	\centering
 	\includegraphics[width=0.8\linewidth]{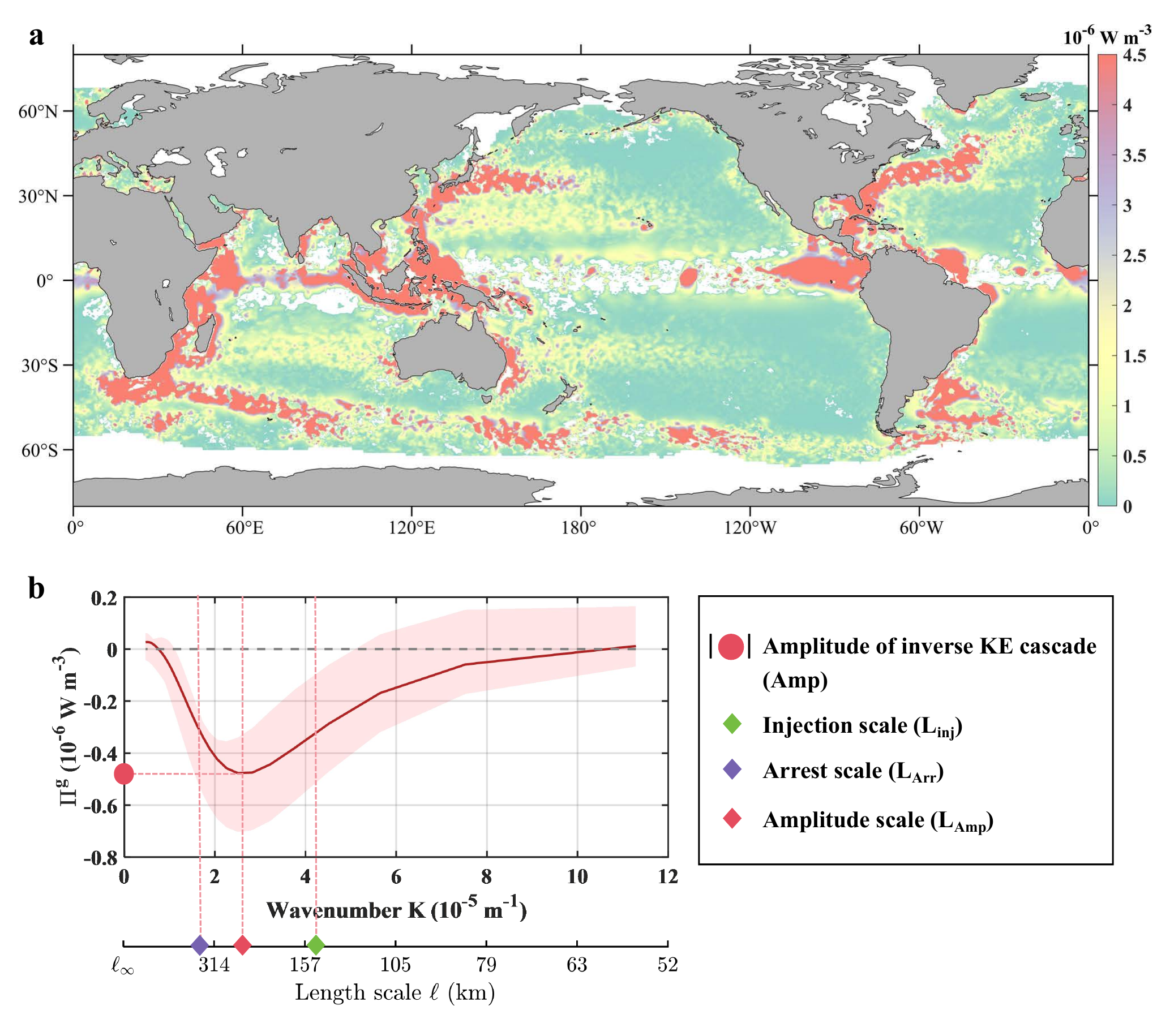}
 	\caption{\textbf{Metrics characterizing  inverse KE cascade}. \textbf{a}, Global map of the climatological amplitude of the inverse cascade from  1994 to 2017. \textbf{b}, Definitions of the  characteristics of the inverse KE cascade. The dark red curve  represents  the globally averaged climatological spectral KE flux ($\varPi ^{g}\left( \ell \right)$), with shading indicating the temporal variability over the period 1994-2017.}
 	\label{fig1}
 \end{figure}

 \section*{Intensified inverse KE cascade \textcolor{black}{in energetic oceans} during 1994-2017}
\begin{figure}[ht]
	\centering
	\includegraphics[width=0.98\linewidth]{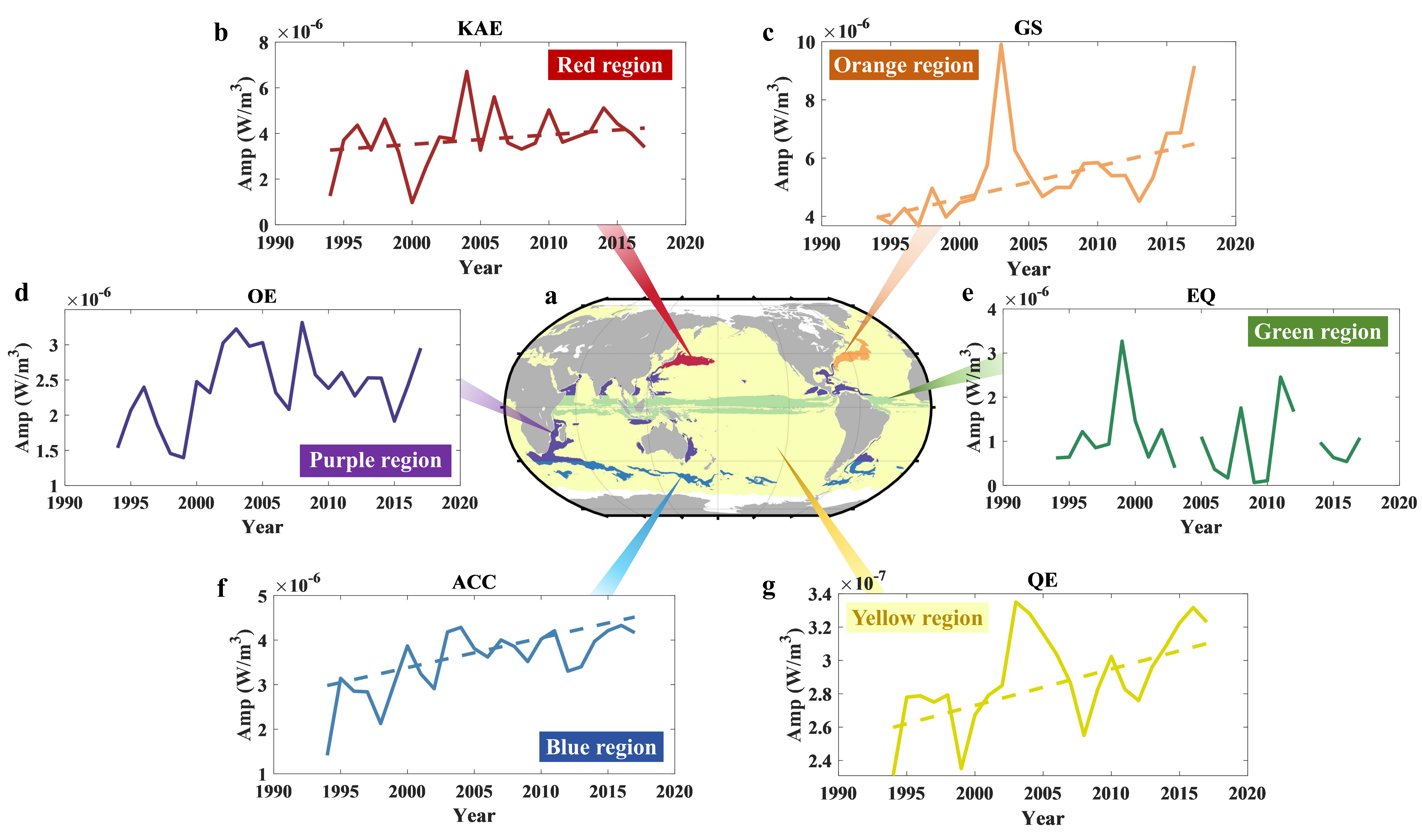}
	\caption{\textcolor{black}{\textbf{Regional trends in the  inverse KE cascade amplitude}. \textbf{a},  Map of ocean regions, highlighting both energetic regions, where TKE ranks in the top 15\% globally, and quiescent regions (QE, yellow mask). The five energetic regions are the Kuroshio currents And its Extension (KAE, red mask), Gulf Stream and its extension (GS, orange mask), EQuatorial region (EQ, green mask), Antarctic Circumpolar Current (ACC, blue mask) and Other Energetic regions (OE, purple mask). \textbf{b}-\textbf{f}, Time series of the regionally averaged inverse KE cascade amplitude ($Amp$) for each region. Dashed lines indicate significant linear trends at the 95\% confidence level. Significant $Amp$ trends are observed in all regions except the  OE (d) and EQ (e) regions.}}
	\label{fig2}
\end{figure}

The trends of regionally averaged $Amp$ reveal a robust intensification of inverse KE cascade across the global ocean from 1994 to 2017 (Fig. \ref{fig2}). This intensification spans not only the most energetic ocean regions, but also extends across vast areas of quiescent regions. A further explanation of "energetic ocean regions" is provided in section `Definition of energetic oceans' in Methods. Given that these energetic ocean regions have significant climate impacts \cite{hu2015pacific,cai2023southern,kelly2010western,kubota2014larger} and exhibit prominent inverse KE cascades (Fig. \ref{fig1}a), we divide the global ocean into five high-energy regions. These regions are categorized based on their distinct dynamical characteristics and geographical location: the Kuroshio and its Extension (KAE), the Gulf Stream and its Extension (GS), Equatorial regions (EQ), the Antarctic Circumpolar Current (ACC), and Other Energetic regions (OE). The remaining areas of the global ocean, characterized by relatively low KE, are collectively defined as the Quiescent regions (QE).

Among the five energetic ocean regions, the inverse KE cascade has significantly intensified during 1994-2017 in the KAE, GS and ACC regions (Fig. \ref{fig2}), with $Amp$ trends increasing at rates of 1.1$\%$, 2.0$\%$ and 1.9$\%$ per decade, respectively (Table \ref{table1}). The OE region also exhibits a strengthening trend, although it is insignificant at the 95$\%$ confidence level (Table \ref{table1}), which is potentially attributed to offsetting effects among subregions exhibiting opposite trends (not shown). In contrast, the EQ region shows an insignificant weakening of the inverse cascade (Table \ref{table1}). The observed $Amp$ trends above are consistent with the recently documented EKE trends \cite{Josu2021Global}, indicating a strong connection between inverse KE cascade and mesoscale activities. In addition to the energetic regions, $Amp$ trends in QE also demonstrate a significant increase at a rate of 0.8$\%$ per decade (Fig. \ref{fig2}g, Table \ref{table1}). This suggests that the intensification of the inverse cascade is widespread throughout the global ocean given the vast expanse of the QE region. \textcolor{black}{Despite the relatively large fluctuations in the $Amp$ time series within these regions, the estimated $Amp$ trends remain robust owing to the effectiveness of the trend analysis method employed here (see `Trend analysis method' in Methods).} The heterogeneity of $Amp$ trend magnitude reflects the high complexity of the processes modulating inverse KE cascade strength in the ocean. 

In addition to $Amp$, three other metrics characterize the inverse KE cascade: the injection scale ($L_{inj}$), the arrest scale ($L_{arr}$) and the amplitude scale ($L_{Amp}$) (definitions available in Fig. \ref{fig1}b and `Metrics of KE cascade characteristics' in Methods). $L_{inj}$ represents the scale where the largest energy source occurs and where the forcing term injects the most KE into the region, whereas $L_{arr}$ corresponds to the scale where energy dissipation takes place, leading to the strongest arrest of upscale KE transfer \cite{li2021barotropic}. $L_{Amp}$ is the scale where the strongest upscale KE transfer occurs and marks the onset of the arrest of the inverse KE cascade \cite{wang2015geographical}. Compared to $Amp$, these three scales ($L_{inj}$, $L_{arr}$ and $L_{Amp}$) exhibit weaker responses to global warming during 1994-2017 (Table \ref{table1}). Only $L_{Amp}$ in the KAE region shows a significantly negative trend.

\begin{table}[htbp]
	 \captionsetup{justification=raggedright, singlelinecheck=false}
	\caption{\textbf{Trends in regional inverse cascade.}}\label{table1}
	\begin{threeparttable}
%	\centering
	\begin{adjustbox}{width=\textwidth}
		\begin{tabular}{|c|c|c|c|c|c|c|}
			\hline
			& \textbf{KAE} & \textbf{GS} & \textbf{EQ} & \textbf{ACC} & \textbf{OE} & \textbf{QE} \\ \hline
			\textbf{$Amp$ Trend ($10^{-8}$ W m$^{-3}$ dec$^{-1}$)} & 4.2 {[}-1.3$\sim$12.3{]} ($\checkmark$) & 11.0 {[}6.3$\sim$15.7{]} ($\checkmark$) & -1.4 {[}-5.6$\sim$4.3{]} & 6.7 {[}3.1$\sim$10.0{]} ($\checkmark$) & 2.6 {[}-0.8$\sim$5.9{]} & 0.2 {[}0.1$\sim$0.4{]} ($\checkmark$) \\ \hline
			\textbf{$L_{inj}$ Trend (km dec$^{-1}$)} & 0 & 0 & 0 & 0 & 0 & 0 \\ \hline
			\textbf{$L_{arr}$ Trend (km dec$^{-1}$)} & 0 & 0 & 6.0 {[}-8.9$\sim$23.4{]} & 0 & 0 & 0 \\ \hline
			\textbf{$L_{Amp}$ Trend (km dec$^{-1}$)} & -1.2 {[}-2.1$\sim$0{]} ($\checkmark$) & 0 & 2.3 {[}-2.4$\sim$9.3{]} & 0 & 0 & 0 \\ \hline
%\bottomrule
\multicolumn{7}{l}{The metrics ($Amp$, $L_{inj}$, $L_{arr}$ and $L_{Amp}$) are defined in Fig. \ref{fig1}b. The uncertainty range represents the 95\% confidence interval. Significant trends are indicated with a “$\checkmark$”.}

		\end{tabular}
	\end{adjustbox}

\end{threeparttable}
\end{table}

\section*{Does change of shear and mixed layer intensify inverse KE cascade? }

Recent observations reveal significant global changes in upper-ocean stratification and mixed-layer depth over the past half-century, characterized by strengthened stratification and deeper mixed layers\cite{li2020increasing,sallee2021summertime}. Under global warming, ocean kinetic energy is projected to increase in the upper ocean\cite{peng2022surface} while decreasing in the deeper ocean \cite{wang2024more}, resulting in an enhanced vertical shear of ocean circulation (shear henceforth). \textcolor{black}{In the context of a two-layer quasigeostrophic (QG) model, enhanced stratification is associated with the increased first baroclinic deformation radius ($R_d$), a deepened mixed layer implies a larger layer thickness ratio ($\delta$), and intensified shear corresponds to a greater horizontal velocity difference between the two layers ($U_1 - U_2$). Idealized QG experiments have demonstrated that variations in $R_d$, $\delta$ and $U_1-U_2$ can significantly modulate the inverse KE cascade\cite{arbic2000generation, 2006Atmospheric, scott2007spectral,smith2001scales}. We therefore hypothesize that the observed changes in stratification, mixed layer depth and shear may contribute to the significant trends in inverse cascade amplitude observed in the energetic regions.}

\textcolor{black}{In this study, we employ a two-layer QG model to formulate a test to our hypothesis. This model has been widely used to investigate dynamical mechanisms of oceanic eddies and jets \cite{panetta1993zonal,smith2001scales,larichev1995eddy,held1996scaling,arbic2000generation,scott2007spectral,arbic2012nonlinear,arbic2014geostrophic,thompson2010jet,thompson2011low,berloff2011latency,chen2016time,thompson2007two}. For example, it has proven effective in simulating jet flow\cite{panetta1993zonal,thompson2010jet}. The jet drift captured in this model was found to be consistent with realistic numerical model results in the Southern Ocean\cite{thompson2011low} (see `A two-layer QG model and the corresponding KE cascade' in Methods). }Using this model, we examine \textcolor{black}{the impact of the following key factors on the inverse cascade amplitude}: deformation radius ($R_d$), layer depth ratio ($\delta$), and the horizontal velocity difference between the upper and lower layers ($U_1-U_2$) (refer to `A \textcolor{black}{two-layer} QG model and the corresponding KE cascade' in Methods for detailed definitions of these factors).  \textcolor{black}{Our general experimental results indicate that, when holding other variables constant, $Amp$ is monotonically increasing with respect to $U_1-U_2$ or $\delta$ individually. However, the relationship between $R_d$ and the inverse cascade amplitude is non-monotonic (Extended Data Fig. \ref{extendfig1}). To further elucidate these dynamics, we conduct targeted sensitivity experiments in regions where inverse KE cascades have intensified, assessing the relative contributions of these three factors to the observed cascade trends.}

\begin{figure}[!ht]
	\centering
	\includegraphics[width=1.0\linewidth]{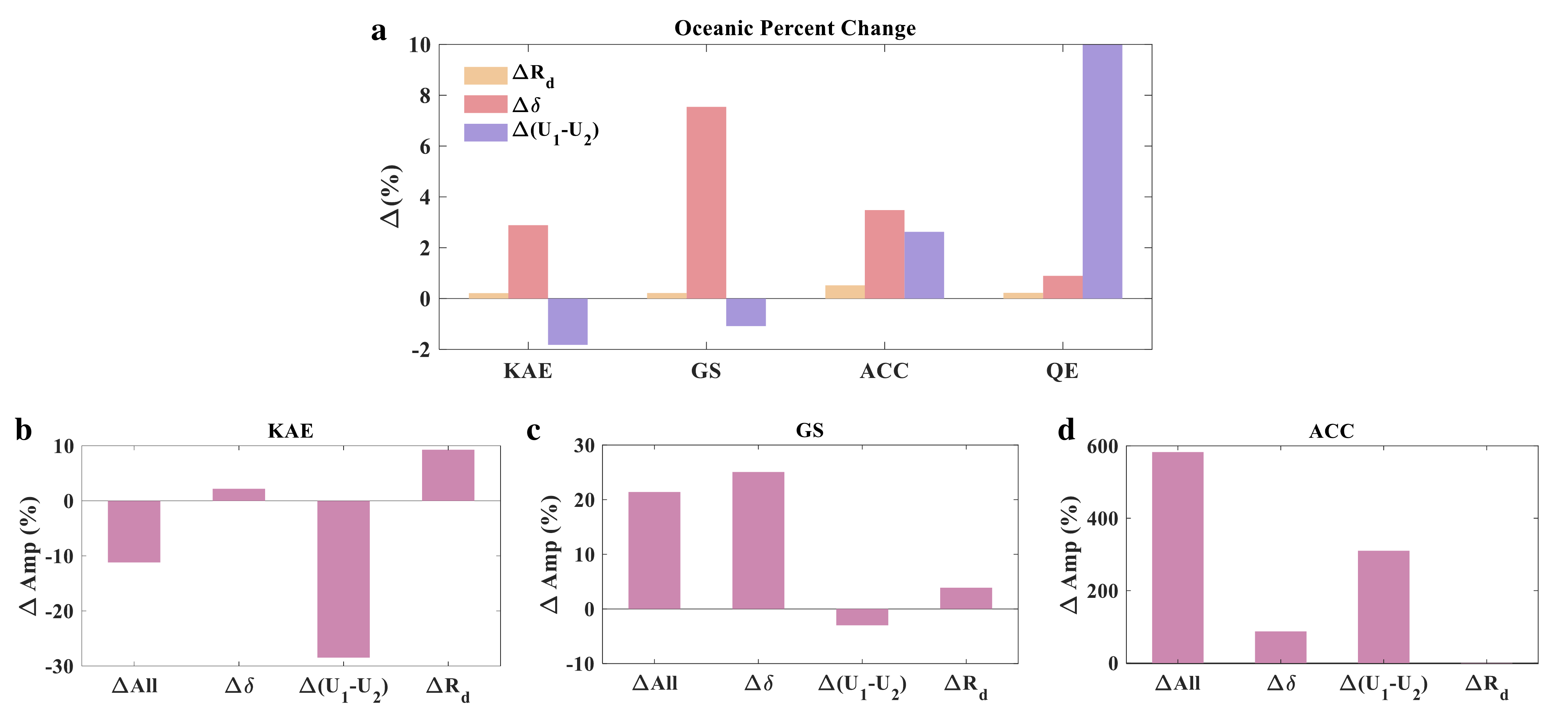}
		\caption{\textbf{Inverse KE cascade response to oceanic changes.} \textbf{a}, Percent changes in regionally averaged $R_d$, $\delta$, and $U_1-U_2$ during 1994-2017, estimated from the ARMOR3D dataset. For the definition of ``percent change'', see `Quantification of oceanic changes' in the Methods section. \textbf{b}-\textbf{d},  \textcolor{black}{Percent change in $Amp$ from the QG experiments, corresponding to the parameter changes shown in (a). The scenarios for the QG sensitive experiments are as follows: [I] original state (1994-2005); [II] overall adjustments of $R_d$, $\delta$, and $U_1 - U_2$ in the later state (2006–2017) relative to the original state [corresponding to the ``$\Delta All$'' experiment]; [III] adjustment of  $\delta$ alone [``$\Delta \delta$'']; [IV] adjustment of  $U_1-U_2$ alone [``$\Delta (U_1-U_2)$''];  and [V] adjustment of  $R_d$ alone [``$\Delta R_d$''].}}
	\label{fig3}
\end{figure}

\textcolor{black}{To design the QG model sensitivity  experiments tailored to the regions we consider, we firstly quantify the observed changes in the above three factors ($R_d$, $\delta$, and $U_1-U_2$) in the corresponding regions (see `Quantification of oceanic changes' in Methods for the metric of oceanic changes).} These factors, closely linked to stratification, mixed-layer depth and shear, are expected to exhibit increasing trends. Analysis of the observation dataset, ARMOR3D, confirms significant trends in these three factors during 1994-2017 (Fig. \ref{fig3}a). In all the four regions with significant $Amp$ trends, i.e., the KAE, GS, ACC and QE regions, the deformation radius ($R_d$) \textcolor{black}{during 2006-2017} has shown a slight increase at a rate of O(0.1)$\%$ \textcolor{black}{compared to 1994-2005}. This aligns with the reported enhancements in stratification.  The layer-depth ratio ($\delta$) has increased more prominently, with rates ranging from 1$\%$ to 7$\%$, consistent with the observed deepening of the mixed layer. Regarding the horizontal velocity difference, $U_1-U_2$, a significant increasing trend is evident in the ACC and QE regions. Conversely, in the KAE and GS regions,  $U_1-U_2$ exhibits a decreasing trend. \textcolor{black}{Previous studies have also identified the decreased vertical shear of zonal velocity in the GS region (Fig. 2F in ref. \cite{peng2022surface}). Yet the underlying mechanisms for the reduced $U_1-U_2$ in both the KAE and GS regions warrant further exploration}. 

Building on the results described above, we conducted five sensitivity experiments using the two-layer QG model for each energetic region with a significant increasing $Amp$ trend (see details of the QG  experiments in `A \textcolor{black}{two-layer} QG model and the corresponding KE cascade' in Methods and Extended Data Table \ref{table2}). The QG model results reveal that the factors driving the cascade intensification vary across regions.  \textcolor{black}{In the GS region, the increased $\delta$ and $R_d$ respectively contribute to a 25\% and 3\% increase in $Amp$, while the reduced $U_1-U_2$ leads to a 2\% decrease (Fig. \ref{fig3}c). This is consistent with our general experiment results (Extended Data Fig. \ref{extendfig1}). The combined variation of the three factors causes a 20\% rise in $Amp$. These results suggest that among the three factors, the increased layer depth ratio ($\delta$), which corresponds to mixed layer deepening, is the dominant factor driving the strengthened inverse cascade in the GS region. In the ACC region, when all the three factors vary concurrently, $Amp$ increases nearly sixfold (Fig. \ref{fig3}d). Again aligning with the general experimental results (Extended Data Fig. \ref{extendfig1}), both the increased $\delta$ and increased $U_1-U_2$ lead to a larger $Amp$. The strengthened flow shear emerges as the primary driver of the increased $Amp$, while the increased layer depth ratio serves as a secondary driver among the three factors. In the KAE region, the sensitivity experiment results are also consistent with the general experiment results. Both the increased $\delta$ and increased $R_d$ correspond to an enhanced $Amp$, while the weakened $U_1-U_2$ correspond to a decreased $Amp$  (Fig. \ref{fig3}b). However, when these factors—$\delta$, $R_d$, and $U_1 - U_2$—are varied simultaneously, their combined effect results in a reduction of $Amp$. This indicates that the QG model employed in this study is still overly idealized. In particular, it does not account for horizontal shear of velocity, which is a key indicator of barotropic instability, and therefore cannot adequately represent this process\cite{waterman2011eddy}. Moreover, due to the use of a doubly periodic numerical domain, the model cannot capture the effects of spatial energy transport. To address this limitation, we further examine the trends in inverse cascade through analysis of the KE budget.}

\section*{\textcolor{black}{Does factors from KE budget  intensify inverse KE cascade?}}
Here we utilize the KE budget to investigate factors consistent with the intensification of inverse KE cascade (see `KE budget analysis' in Methods). Our analysis mainly focuses on six terms within the KE budget at the separation scale $L_{Amp}$: the geostrophic KE cascade term ($\varPi^{g}$), the KE transport term by geostrophic flow ($J^{g,flow}$), the geostrophic component of the pressure transport term  ($J^{g,pre}$), the tendency term of small-scale geostrophic KE ($\frac{\partial KE_{S}^{g}}{\partial t}$), the geostrophic component of wind power input term ($WP^g$) and the residual terms ($OT$) (detailed interpretations are provided in `KE budget analysis' in Methods and Extended Data Table \ref{table3}). The net changes in these terms are linked to the change in $Amp$, \textcolor{black}{potentially through modulating  the amount of KE available for inverse cascade processes (see `Change of KE budget' in Methods).} 

\begin{figure}[!ht]
	\centering
	\includegraphics[width=0.9\linewidth]{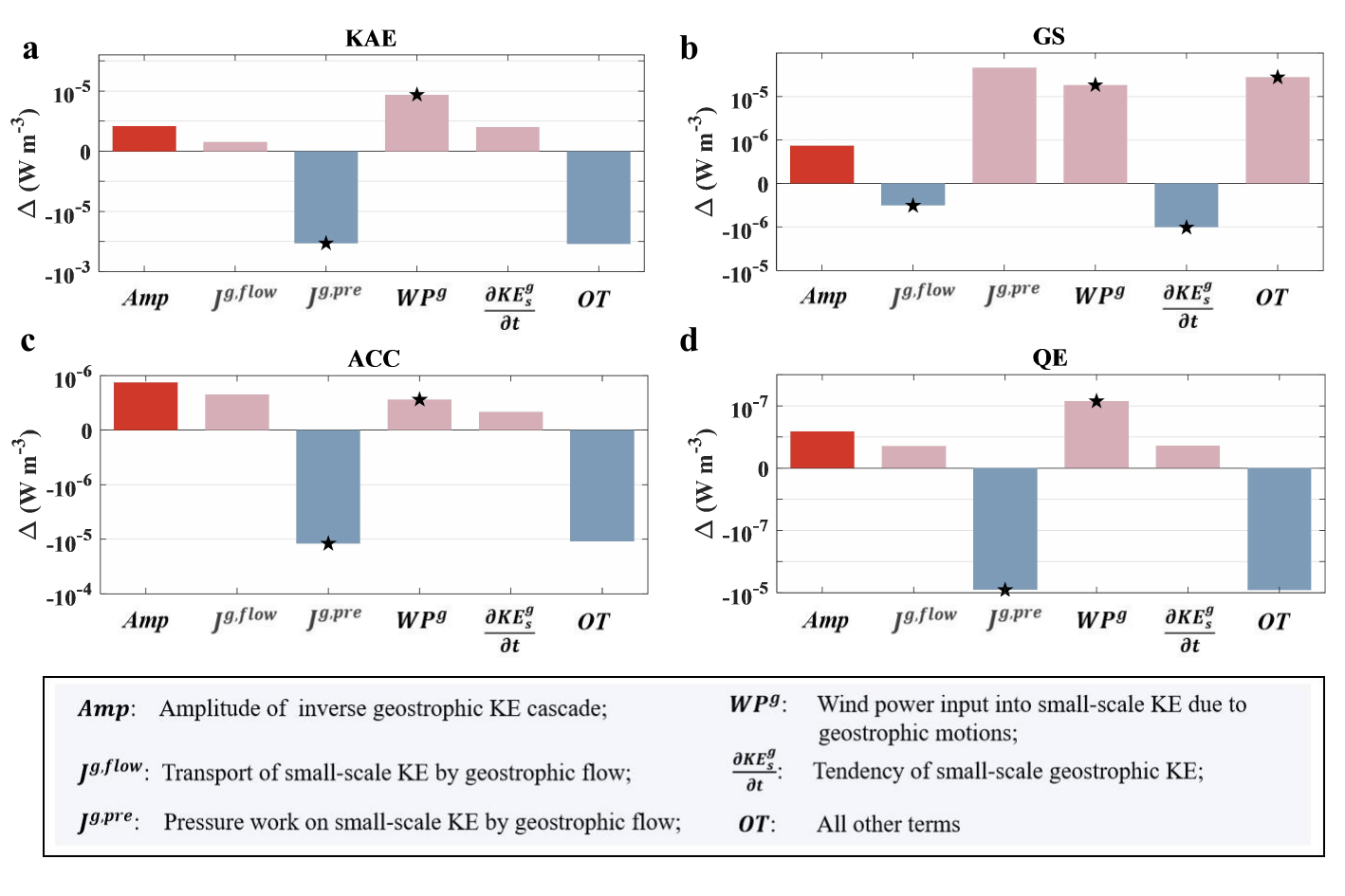}
	\caption{\textcolor{black}{\textbf{Changes in the regional KE budget.} \textbf{a}-\textbf{d}, Differences in the small-scale KE budget for each region between 2006-2017 and 1994-2005, at the separation scale $L_{Amp}$.  An increasing (decreasing) budget term is represented by a pink (blue) bar. Terms contributing to KE cascade intensification  are marked with a black “$\star$”. For details, refer to `KE budget analysis', `Change of KE budget' in the Methods section, Extended Data Table \ref{table3}, and Supplementary Information.}} 
	\label{fig4}
\end{figure}

In the KAE, ACC, and QE regions, the budget terms associated with the intensified inverse cascade are consistent (marked terms in Fig. \ref{fig4}). Both the pressure transport term ($J^{g,pre}$) and the wind power input term ($WP^g$) contribute positively to the intensification of the KE cascade.  A magnitude comparison of these changes indicates that the reduction in $J^{g,pre}$ plays a more significant role in cascade intensification than the changes in $WP^g$. In contrast, the GS region exhibits distinct characteristics. All the terms, except for the pressure transport term ($J^{g,pre}$), contribute to the enhanced inverse cascade observed in this region. \textcolor{black}{
Among these four regions, the magnitudes of $OT$ term are comparable to those of $J^{g,pre}$. Nevertheless, in the KAE, ACC and QE regions, the $OT$ term does not contribute positively to the intensification of inverse KE cascade. In the GS region, despite the $OT$ term appears to support the intensified inverse cascade, a detailed interpretation of its role is not pursued here, due to the lack of data concerning the various processes encapsulated within this term (refer to Extended Data Table \ref{table4}). Therefore, although uncertainties are contained in the $OT$ term, the results here remain robust}. 

\section*{Does wind-induced eddy killing intensify inverse KE cascade?}
The KE budget analysis reveals that in all the four regions with intensified inverse cascade, the increased wind power input ($WP^g$) in the KE budget positively contributes to the inverse cascade enhancement (Fig. \ref{fig4}). This increased wind power input into small-scale KE ($WP^g$) \textcolor{black}{corresponds to the weakening of  "eddy killing". Eddy killing refers to the damping effect of atmospheric winds on oceanic eddies \cite{dewar1987some,duhaut2006wind,dawe2006effect,zhai2007wind,hughes2008wind,zhai2012wind,xu2016work,wilson2016does,seo2016eddy,renault2016control,wilder2022response,rai2021scale}. When eddy killing occurs, the wind power input into eddies is negative,  acting as a key energy sink for mesoscale currents.} \textcolor{black}{Eddy killing is based on the premise that wind stress is a function of the relative air-sea velocities, a phenomenon known as the "relative wind stress effect". Since the spatial scales of atmospheric winds generally exceed those of oceanic eddies, the background wind field is typically assumed to be spatially uniform. This simplification enables a focus on the vortex structures of the eddies when investigating their interactions with the wind.} Consider an anticyclonic eddy in the Northern Hemisphere under a prevailing westerly wind. As the wind blows over the eddy, the wind-eddy interaction would produce a smaller wind stress on the northern side of the eddy, where the wind and currents are aligned; and a larger wind stress on the southern side where they are in the opposite direction. This differential wind stress generates greater negative wind work on the southern half of the eddy compared to the positive wind work on the northern half. When integrated spatially, this imbalance results in a net negative wind power input, which consequently extracts energy from the eddy and leads to "eddy killing".

\textcolor{black}{ \textcolor{black}{As stated in literature\cite{rai2021scale}, a negative} value for $WP^g$ indicates the occurrence of eddy killing. Our analysis shows that at the spatial scale $L_{Amp}$ (mainly within mesoscale range), the climatological $WP^g$ (i.e., the time mean during 1994-2005) is negative in \textcolor{black}{all the four regions}  (Extended Data Fig. \ref{extendfig2}a), suggesting a robust occurrence of eddy killing. \textcolor{black}{Compared to the 1994–2005 period, the time-mean value of $WP^g$ during 2006–2017 has significantly increased, indicating a weakening of eddy killing across these regions (Extended Data Fig. \ref{extendfig2}b).} Given the recently discovered synchronous variation between the inverse KE cascade and KE spectrum in the mesoscale range \cite{storer2023global}, the reduced eddy killing could result in an increase in the KE available for the inverse KE cascade, thereby enhancing the rate of upscale KE transfer. Notably, the magnitude of the increase in $WP^g$, \textcolor{black}{ranging from $O(10^{-7})$ $W/m^3$ to $O(10^{-5})$ $W/m^3$,} is comparable to the increase in $Amp$ in the ACC region and even up to ten times larger than the $Amp$ change observed in the KAE, GS and QE regions. This highlights the undeniable role of eddy killing in intensifying inverse KE cascade.}

\section*{Discussion}
\begin{figure}[!ht]
	\centering
	\includegraphics[width=0.95\linewidth]{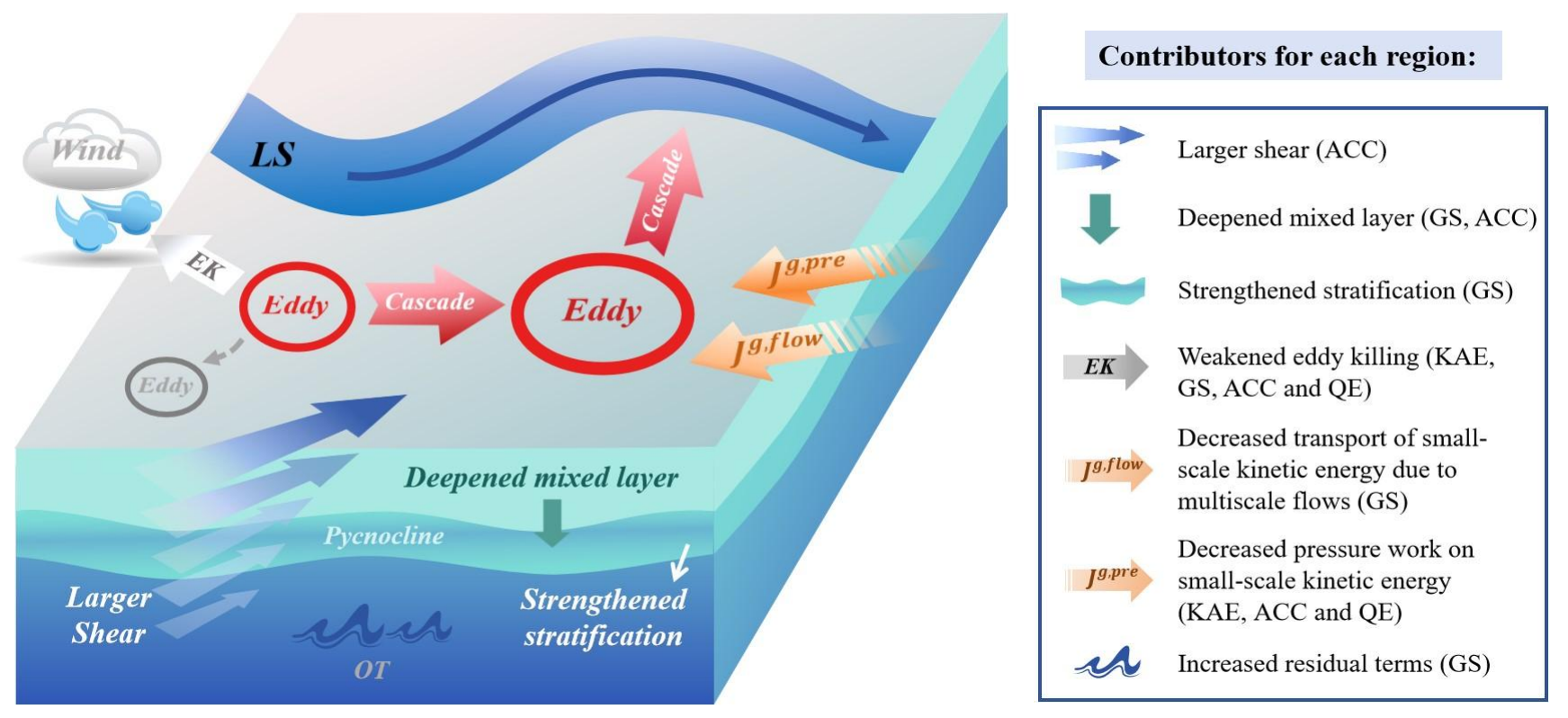}
	\caption{ \textcolor{black}{\textbf{Schematic diagram of the inverse KE cascade intensification and its underlying mechanism.} The abbreviation LS refers to large-scale flows. The two red arrows represent the inverse KE cascades, which include cascades from relatively small to relatively large eddies and from eddies to large-scale flows. Other components in the diagram illustrate the key contributors to KE cascade intensification, as further explained in the box on the right. For instance, in the KAE region, both weakened eddy killing and reduced pressure work contribute to the enhanced inverse KE cascade. In the ACC region, increased vertical shear and mixed layer depth, along with weakened eddy killing and decreased pressure work, all contribute to KE cascade intensification. For further details on the KE budget terms shown here, refer to the “KE budget analysis” in the Methods section and Extended Data Table \ref{table3}.}}
	\label{fig5}
\end{figure}

Based on available satellite observations, we have found evidence of an overall intensification of the surface inverse KE cascade \textcolor{black}{amplitude} across the global ocean during 1994-2017. \textcolor{black}{The regions with intensified cascade amplitude include both the energetic ocean areas  and the quiescent regions.} The injection, arrest and amplitude scales for inverse KE cascade, however, show little net change. The factors consistent with the observed trends in inverse KE cascade show multifaceted complexity and large regional heterogeneity (Fig. \ref{fig5}). In both the Kuroshio and quiescent regions, the weakening of eddy killing and the decrease in pressure work ($J^{g,pre}$) contribute to the enhancement of the inverse cascade. \textcolor{black}{In the ACC region, alongside changes in eddy killing and pressure work, increased vertical shear of horizontal velocity and a deepened mixed layer also play key roles. The Gulf Stream region presents a more intricate scenario. Here, a deepened mixed layer, strengthened stratification, reduced transport from multiscale flows, weakened eddy killing, and changes in the residual terms within the KE budget all contribute positively to the inverse KE cascade.} 

A potential linkage can be identified between the results from the QG model and the KE budget. In the QG dynamical regimes we consider, the increased mixed layer depth, stratification and vertical shear of horizontal velocity could enhance the KE inverse cascade, which is partially attributed to strengthened baroclinic instability and thus mesoscale eddy activity. In the KE budget framework, information about baroclinic instability is contained in the $OT$ term (see `KE budget analysis' in Methods and Extended Data Table \ref{table3}). \textcolor{black}{
	Due to the lack of vertical velocity data in current observational datasets, directly quantifying the contribution of baroclinic instability in the KE budget remains challenging. Nevertheless, the results from QG model and KE budget in this study offer valuable insights into the mechanism of the intensified inverse KE cascade. }

\textcolor{black}{There are several potential avenues for future research. Firstly, the trend analysis in this study could benefit from the use of model data with longer time series or higher resolutions, such as those provided by the high-resolution CMIP (Coupled Model Intercomparison Project) simulations. The advantage of model data lies in its extended temporal coverage, which enables independent validation of observed trends and facilitates the separation of trend signals from long-timescale oscillations. \cite{huybers2005obliquity}. However, model simulations are often subject to systematic biases such as model drift\cite{jin2009characteristics,vanniere2013using,reed2015analysis}. In contrast, satellite observations offer more reliable estimates at resolved spatial scales, but their temporal coverage is relatively limited, and they are unable to resolve submesoscale or smaller scale processes. This limitation may lead to biases in estimating the strength of the KE cascade\cite{qiu2014seasonal}. In the future, a combined approach integrating satellite observations with high-resolution numerical models may allow for a more robust assessment of the cascade trend and its underlying mechanisms. Besides, whether our results here are contaminated by the increasing number of satellites also needs further exploration\cite{barcelo2024robust,o2015global,Josu2021Global}. In addition, given that our results are confined to the surface ocean, a mass of available high-resolution in situ observations underwater, especially in the abyssal ocean, are urgently needed for a full-depth assessment of the KE cascade evolution.}

The intensification of inverse KE cascade we have observed has potentially significant climate implications. Our work provides support for recent studies reporting the global acceleration of surface large-scale circulation \cite{2020Deep,peng2022surface}. In most oceanic regions, the intensified inverse KE cascade indicates a stronger transfer of KE from the mesoscale to larger scales (Fig. \ref{fig1}b and Fig. \ref{fig2}), such as the gyre-scale at which the large-scale circulation dominates. Although being a mere fraction of energy source compared to other sources such as \textcolor{black}{the wind power input \cite{zhai2012wind,hughes2008wind,wunsch2004vertical}}, this increased mesoscale-to-gyre-scale KE transfer can lead to a considerable speed-up of the gyre-scale ocean circulation in the future, owning to the spatial consistency and long-term stability of gyre scales \cite{storer2023global}. In addition,  \textcolor{black}{since mesoscale eddies  generally  originate  from the large-scale circulation through baroclinic and barotropic instabilities} \cite{wunsch2004vertical,qiu2022bi}, the acceleration of ocean circulation could foster more eddy activities, \textcolor{black}{promoting the recently observed EKE strengthening \cite{Josu2021Global}.} This, in turn,  would further enhance the upscale KE transfer from mesoscale to larger scales. Therefore, the amplification of inverse KE cascade contributes to  a feedback cycle involving changes in both large-scale circulation and mesoscale currents. These oceanic changes, partially induced by climate change,  would, in turn, feed back on the Earth’s climatic system. \textcolor{black}{For instance, alterations in the western boundary currents could affect their capacity to transport heat from lower to higher latitudes, thereby influencing the global climate \cite{oceanrole2018,wu2012enhanced,hu2015pacific}; Changes in wind-driven circulation and mesoscale eddy mixing could impact the ocean’s carbon sink\cite{ito2015sustained,southernocean2024,lovenduski2008toward,le2007saturation, gnanadesikan2015isopycnal} , further altering atmospheric carbon levels.}

\section*{Methods}
\subsection*{Observation products}
Multiple datasets are used in this study, all of which have global coverage and span a 24-year period (1994/01/01-2017/12/31). To quantify the trend in the KE cascade, we estimate the spectral KE flux using the geostrophic current velocities from the Archiving, Validation, and Interpretation of Satellite Oceanographic Data (AVISO) provided by the French National Space Agency, also accessible through the Copernicus Marine and Environment Monitoring Service (CMEMS). The horizontal resolution of this gridded dataset is $1/4^{\circ}$, with a temporal resolution of one day.

\textcolor{black}{For the mechanism analysis, we utilize version 3.1 of the CCMP (Cross-Calibrated Multi-Platform) global wind speed dataset \cite{mears2022improving} from Remote Sensing Systems to diagnose \textcolor{black}{ the wind power input term in the KE budget}. This dataset integrates sea surface wind speed data from multiple satellite microwave sensors using a Variational Analysis Method, with reanalysis data as the background field \cite{mears2022improving}. It provides 6-hourly vector wind at a spatial resolution of $1/4^{\circ}$.}

To quantify the changes in shear, stratification and mixed layer depth, we use the ARMOR3D dataset from CMEMS. This global three-dimensional observational dataset integrates various in situ observations—including Argo profilers, CTDs, XBTs, and moorings—with satellite altimetry data. \textcolor{black}{ It  provides $1/4^{\circ}$ gridded data on ocean temperature, salinity, potential height, geostrophic currents from the surface to  approximately 5500 m depth, as well as two-dimensional mixed layer depth. }
\subsection*{Quantification of KE cascade}
KE cascade is analyzed through the spectral KE flux $\varPi \left( x,y;t;\ell \right)$ (Fig. \ref{fig1}b), which measures the KE transfer rate from scales larger than $\ell$ to smaller scales. In this study, we employ a filter-based coarse-graining approach \cite{Aluie2018Mapping} for calculation \textcolor{black}{of the spectral KE flux. We focus on its geostrophic component, }
\begin{equation}
	\varPi^{g}\left( x,y;t;\ell \right) =-\rho _0\left[ \tau _{\ell}\left( u^g,u^g \right) \frac{\partial u_{\ell}^{g}}{\partial x}+\tau _{\ell}\left( u^g,v^g \right) \left( \frac{\partial u_{\ell}^{g}}{\partial y}+\frac{\partial v_{\ell}^{g}}{\partial x} \right) +\tau _{\ell}\left( v^g,v^g \right) \frac{\partial v_{\ell}^{g}}{\partial y} \right], 
	\label{eq1}
\end{equation}
where $u$ and $v$ denote the zonal and meridional current velocities, respectively, and the superscript “g” denotes geostrophic variables. The subfilter stress tensor, 
\begin{equation}
	\tau _{\ell}\left( u^g,u^g \right) =\left( u^gu^g \right) _{\ell}-{u}_{\ell}^g{u}_{\ell}^g,
	\label{eq12}
\end{equation}
\textcolor{black}{quantifies the forces exerted by small-scale motions (at scales <$\ell$) on large-scale motions (at scales >$\ell$)}\cite{Aluie2018Mapping}. $\rho _0$ is seawater referenced density and takes a constant of 1027.4 $kg\cdot m^{-3}$. The filter approach follows ref.\cite{rai2021scale}.

Using the coarse-graining approach, we can quantify the energy transfer rate at any space point $\left( x,y \right)$ and any instant of time $t$ \cite{Aluie2018Mapping}. The negative (positive) values of $\varPi ^{g}$ indicate the inverse (forward) KE cascade. 

\subsection*{Metrics of KE cascade characteristics}
We consider four metrics of KE cascade characteristics (Fig. \ref{fig1}b): inverse KE cascade amplitude ($Amp$), injection scale ($L_{inj}$), arrest scale ($L_{arr}$) and amplitude scale ($L_{Amp}$). $Amp$ is the maximum value of $-\varPi$ over the wavenumber range where the inverse KE cascade occurs \cite{wang2015geographical}.  Therefore, the value of $Amp$ is always positive. A larger $Amp$ value represents a stronger inverse KE cascade, that is, more KE is transferred to larger scales. The variables $L_{inj}$ and $L_{arr}$ \textcolor{black}{correspond to the scales where the} spectral flux exhibits the steepest positive and negative slopes, respectively \cite{li2021barotropic}. $L_{inj}$ corresponds to the largest energy source, while $L_{arr}$ corresponds to the largest energy sink \cite{li2021barotropic}. $L_{Amp}$ is defined as the scale where the spectral KE flux reaches its $Amp$ value, and can be considered the scale at which upscale energy transfer begins to be arrested \cite{wang2015geographical}.  

In literature \cite{2005Direct,li2021barotropic,wang2015geographical,qiu2008length,tulloch2011scales}, the injection scale  $L_{inj}$ is often defined as the spatial scale where the spectral KE flux is zero and has a positive slope over a range of wavenumbers. It also represents the start of inverse cascade. Here we choose to use a different definition of $L_{inj}$ for two reasons: \textcolor{black}{ i) The definition of $L_{inj}$ in this study has a more intuitive physical interpretation compared to the classic definition. Physically, $L_{inj}$ in this study represents the scale at which energy injection is strongest, which aligns well with the definition of $L_{arr}$ in this study. \textcolor{black}{In contrast, the classic definition of the injection scale is less representative, as it indicates neither the start nor the peak of energy injection}; ii)  Previous studies often employed the "spectral approach" \cite{schubert2020submesoscale} to diagnose energy cascade. Using this method, one can easily identify the zero-crossing point of spectral KE flux at large wavenumbers, which corresponds to the classic injection scale.  However, the spectral approach has two significant limitations compared to the coarse-graining method used here: it lacks Galilean invariance and includes a transport term that does not accurately reflect energy cascade dynamics \cite{Spectralkinetic2022}. The coarse-graining method avoids these issues and is therefore more robust. However, a zero-crossing point is generally absent in the spectral flux derived from the coarse-graining approach \cite{storer2023global}. To accommodate the coarse-graining approach, we redefine $L_{inj}$ in this study.}    

 \subsection*{Trend analysis method}
\textcolor{black}{Trends are estimated using the Theil-Sen (TS) estimator method\cite{theil1950rank,sen1968estimates} from the regionally-averaged time series of each spectral flux metric we consider (Fig. \ref{fig2}, \textcolor{black}{Table \ref{table1}}).}  Compared to the conventional least square method, \textcolor{black}{the TS estimator is less sensitive to outliers in the data\cite{ohlson2015linear,chervenkov2019theil}, making it a more effective and robust approach for estimating linear trends. As a result, it has been widely used in meteorological and oceanographic studies \cite{gocic2013analysis,sa2019trends,zhao2022uncertainty}}. For trend significance  \textcolor{black}{assessment, we use} a modified Mann-Kendall test \cite{mann1945nonparametric,hamed1998modified}, which has been widely used in ocean change analysis \cite{2020Deep,Josu2021Global,zhang2023global}. The significance level in this study is set to be 0.05 both for the trend significance (Fig. \ref{fig2}) and the uncertainty ranges (\textcolor{black}{Table \ref{table1}}).

\subsection*{Definition of energetic oceans}
\textcolor{black}{The ``energetic oceans'' in this study are defined as the areas where KE ranks in the top 15\% of all the global grid points (Fig. \ref{fig2}a). These regions are also found to exhibit strong inverse energy cascades (Fig. \ref{fig1}a), which motivates our focus on them.} We divide the whole  ``energetic oceans''  into five distinct areas according to their dynamics and geographical locations: the Kuroshio current \textcolor{black}{And its E}xtension (KAE), Gulf Stream and its extension (GS), EQuatorial regions (EQ), Antarctic Circumpolar Current (ACC) and Other Energetic regions (OE, all the remaining energetic ocean regions). \textcolor{black}{In this study, ``KE'' refers to kinetic energy, \textcolor{black}{whereas KAE refers to the Kuroshio current And its Extension region}.} 

\subsection*{A \textcolor{black}{two-layer} QG model and the corresponding KE cascade}
We simulate the KE cascade response to oceanic changes using a widely-used \textcolor{black}{two-layer} quasigeostrophic (QG) model on a $\beta$ plane  \cite{thompson2010jet,berloff2011latency}.  This model has proven to be effective for studying a variety of oceanic dynamical problems, such as eddies, jets, multiscale interactions and machine-learning parameterizations \cite{panetta1993zonal,smith2001scales,larichev1995eddy,held1996scaling,arbic2000generation, arbic2004baroclinically,thompson2010jet,thompson2011low, arbic2013eddy,chen2016time,berloff2011latency,george2021deep, ross2023benchmarking}. \textcolor{black}{For example, it is demonstrated that this QG model is capable of accurately representing horizontal mesoscale eddies\cite{arbic2004baroclinically}. Additionally, using this QG model, previous studies have successfully simulated zonal jet flows and identified the phenomenon of jet drifting \cite{panetta1993zonal,thompson2010jet}. This finding was later corroborated by a more realistic ocean general circulation model, which confirmed the presence of jet drifting in the Southern Ocean \cite{thompson2011low}. Collectively, these studies highlight the capacity of QG models to realistically capture key dynamical processes observed in the ocean. }
	
Imposed by a horizontally homogeneous and vertically sheared mean flow, this QG model is baroclinically unstable. Its governing equation is ($i$=1 and 2 denotes the upper and lower layer, respectively):
\begin{equation}
	\left( \frac{\partial}{\partial t}+U_i\frac{\partial}{\partial x} \right) q_i+J\left( \psi _i,q_i+\overline{q_i} \right) =ssd-\gamma \delta _{i2}\nabla ^2\psi _2, 
	\label{eq2}
\end{equation}
where $U_i$ represents zonal mean flow, $\psi _i$ represents eddy streamfunction, $\delta _{ij}$ is the Kronecker delta, and $\gamma$ is the bottom Ekman friction coefficient. Small-scale dissipation is represented by $ssd$ and is implemented using a wavenumber cutoff filter \cite{arbic2000generation,chen2016time}. The eddy potential vorticity (PV) $q_i$ and mean PV $\overline{q_i}$ are defined as:
\begin{equation}
	q_i=\nabla ^2\psi _i+\left( -1 \right) ^i\frac{1+\delta _{i2}\left( \delta -1 \right)}{\left( 1+\delta \right) Rd^2}\left( \psi _1-\psi _2 \right) \ and \label{eq3} \\ 
\end{equation}
\begin{equation}
	\overline{q_i}=\beta y+\left( -1 \right) ^i\frac{1+\delta _{i2}\left( \delta -1 \right)}{\left( 1+\delta \right) Rd^2}\left( U_2-U_1 \right) y, \label{eq4}
\end{equation}
respectively. \textcolor{black}{Here $\beta$ is the planetary PV meridional gradient} and $R_d$ is the single deformation radius. The depth ratio $\delta$ is defined as $\delta =H_1/H_2$, where $H_1$ and $H_2$ separately denote the depths of upper-layer and lower-layer.

\textcolor{black}{For each energetic region, we perform five model experiments (see Extended Data Table \ref{table2}).  The parameters corresponding to oceanic changes include \textcolor{black}{the }layer depth ratio ($\delta$), the vertical shear of zonal mean flow ($U_1-U_2$) and \textcolor{black}{the deformation radius ($R_d$)} \textcolor{black}{(for specific values, see Extended Data Table \ref{table2}). \textcolor{black}{These parameter values were obtained by analyzing the observation dataset  ARMOR3D.} The upper layer thickness, $H_1$, is set as the sum of mixed layer depth and half of the pycnocline depth in each region, and $H_1+H_2$ is set to 1 km. $U_i$ is the vertical average of the zonal velocities at each layer.} The numerical domain is} doubly periodic and its size is 1000 km × 1000 km, large enough for the development of inverse KE cascade. \textcolor{black}{Initialized by small-amplitude random motions, the model experiments reach statistical equilibrium when the eddy energy extracted from the mean flow shear through baroclinic instability is balanced by energy dissipation through bottom drag. For each experiment, we analyze 100-year model output in  the statistical equilibrium state.}

\textcolor{black}{KE cascade in this QG model is diagnosed employing the widely used spectral approach \cite{qiu2008length,li2021barotropic,wang2015geographical}. \textcolor{black}{At a specific wavenumber $K ^\prime$, the upper-layer spectral flux $\varPi \left( K ^\prime \right)$ is defined as} \cite{arbic2013eddy}:
\begin{equation}
	\varPi \left( K ^ \prime \right) =\iint\limits_{K >K ^ \prime}{\mathrm{Re}\left( \frac{\delta}{1+\delta}\hat{\psi}_{1}^{*}\widehat{J\left( \psi _1,\nabla ^2\psi _1 \right) } \right) dK}.
	\label{eq5}
\end{equation}
Here \textcolor{black}{$K$ is the isotropic wavenumber}. $\widehat{\cdot }$ denotes performing a Fourier transform on a variable in wavenumber space. $*$ denotes a complex conjugate, and Re represents taking the real part of a complex number. }

\textcolor{black}{Previous studies have conducted excellent investigations on barotropic–baroclinic mode interactions when examining energy cascades using the QG model \cite{larichev1995eddy,arbic2000generation,smith2001scales,arbic2004baroclinically}. They found that the inverse energy cascade is strongly influenced by the energy distribution between the baroclinic and barotropic modes, as well as the efficiency of barotropization. These dynamics can be further modulated by factors such as stratification, the $\beta$ effect, and bottom friction. In this study, we focus specifically on the strength of the KE cascade in the upper ocean layer. As described by ref. \cite{2005Direct}, in a two-layer ocean model, altimetric measurements primarily reflect the motion of the upper ocean layer. The upper-layer streamfunction comprises both barotropic and baroclinic components and can be expressed as:
\begin{equation}
	\psi_{1} = \psi_{\mathrm{bt}} + \frac{\psi_{\mathrm{bc}}}{\sqrt{\delta}},
\end{equation}
where $\psi_1$ denotes the streamfunction of the upper ocean layer, $\psi_{BT}$ is the barotropic component, $\psi_{BC}$ is the baroclinic component, and $\delta$ is the layer thickness ratio. When $\delta$ is particularly small, the contribution from the barotropic mode becomes negligible, and $\psi_1$ is dominated by the baroclinic streamfunction. In our two-layer QG configuration, however, $\delta$ ranges from 0.17 to 0.21 (Extended Data Table \ref{table2}), indicating that the baroclinic contribution to $\psi_1$ is approximately twice that of the barotropic contribution. Therefore, the upper-layer streamfunction represents a mixture of both modes rather than a pure first baroclinic mode. For this reason, our analysis does not focus on the conversion between barotropic and baroclinic modes. }

\textcolor{black}{In addition, as demonstrated in previous studies using the QG model \cite{held1996scaling,gallet2020vortex,gallet2021quantitative}, bottom friction, planetary $\beta$, and topographic $\beta$ can also influence the strength and scale of the inverse KE cascade. However, in our study, both planetary and topographic $\beta$ remain nearly constant across our study regions, and bottom friction cannot be reliably estimated from available observations. Therefore, we do not include the effects of these parameters when interpreting the trends in the inverse KE cascade.}

\subsection*{Quantification of oceanic changes}
\textcolor{black}{In this study, ``oceanic changes'' represent the difference in time-mean variables between 2006–2017 and 1994–2005,} 
\begin{equation}
	\varDelta a=\overline{a}\left| _{}^{2006-2017} \right. -\overline{a}\left| _{}^{1994-2005} \right., 
	\label{eq6}
\end{equation}
\textcolor{black}{where $\varDelta a$ denotes the change of one arbitrary ocean variable (e.g., deformation radius),
 and \textcolor{black}{${\left. {\bar  \cdot } \right|^{Y1 - Y2}}$} denotes temporal averaging over the period from year Y1 to year Y2. The percent change can be calculated as:}
\begin{equation}
	\varDelta a\left( \% \right) =\frac{\varDelta a}{\overline{a}\left| _{}^{1994-2005} \right.}\times 100.
	\label{eq7}
\end{equation}

\subsection*{\textcolor{black}{KE budget analysis}}
\textcolor{black}{Inspired by Equation (\ref{eq12}), we define the small-scale KE as}
	\begin{equation}
		KE_S=\frac{1}{2}\rho _0\tau _{\ell}\left( \mathbf{u},\mathbf{u} \right),
	\end{equation} 
where $\mathbf{u}$ denotes the 3-dimensional current velocity vector. \textcolor{black}{This definition is also consistent with the scale-dependent eddy energy in ref. \cite{steinberg2022seasonality}.} \textcolor{black}{Based on this definition, \textcolor{black}{for a specific filter scale $\ell$,} $KE_S$ represents the difference between total kinetic energy ($KE_T$) and large-scale KE ($KE_L$, \textcolor{black}{i.e., KE contained in oceanic motions  with spatial scales larger than  $\ell$) (refer to Extended Data Table \ref{table4} for mathematical forms of $KE_T$ and $KE_L$).} Therefore, the small-scale KE budget equation can be obtained through subtracting the large-scale KE budget  \cite{Aluie2018Mapping} from the low-pass filtered total KE budget (see detailed derivation in Supplementary Information):}
\begin{equation}
	\begin{aligned}
		\frac{{\partial K{E_S}}}{{\partial t}} =  &- \underbrace {\nabla  \cdot \left[ {{{\left( {{\bf{u}}K{E_T}} \right)}_\ell } - {{\bf{u}}_\ell }K{E_L} - {\rho _0}{\bf{u}}_\ell ^{} \cdot {\tau _\ell }\left( {{\bf{u}},{\bf{u}}} \right)} \right]}_{{J^{flow}}} - \underbrace {\nabla  \cdot {\tau _\ell }\left( {{\bf{u}},P} \right)}_{{J^{pre}}} + \Pi  + \underbrace {\frac{{{{\left( {{\bf{\tau }}_{}^{wind} \cdot {\bf{u}}} \right)}_\ell } - {\bf{\tau }}_{\ell}^{wind} \cdot {{\bf{u}}_\ell }}}{H}}_{WP}
		\\
		&+ {\left( {{\rho _0}{\bf{g}} \cdot {\bf{u}}} \right)_\ell } - {\rho _\ell }{\bf{g}} \cdot {{\bf{u}}_\ell } + {\left( {Dif{f_T}} \right)_\ell } - Dif{f_\ell } + {\rho _0}{\tau _\ell }\left( {{\bf{u}},{\bf{F}}_{}^{forcing}} \right).
		\label{eq13}
	\end{aligned}
\end{equation}
Here, $\nabla$ is the 3-dimensional divergence operator. $P$ denotes pressure. $\tau ^{wind}$ represents wind stress. \textcolor{black}{$H$ approximately corresponds to Ekman layer depth, which typically ranges from O(10) to O(50) meters}\cite{2006Atmospheric,cushman2011introduction}. $g$ is the gravitational acceleration. $\mathbf{F}_{}^{forcing}$ represents the external force vector (mainly tidal forcing). 

\textcolor{black}{Equation (\ref{eq13}) describes the budget of small-scale KE.} $J^{flow}$ represents transport of $KE_S$ due to multiscale flows, while $J^{pre}$ represents pressure work on $KE_S$. \textcolor{black}{ The energy cascade term, $\Pi$, represents the spectral KE flux and is often derived from  the large-scale KE budget in previous literature \cite{Aluie2018Mapping} (see detailed derivation in Supplementary Information). The term $WP$ quantifies the wind power input into ocean motions at scales smaller than $\ell$}\textcolor{black}{, and is linked to eddy killing in literature\cite{rai2021scale}}. A negative value for $WP$ indicates the occurrence of eddy killing\cite{rai2021scale}. For the mathematical forms and physical meanings \textcolor{black}{of the other terms in Equation (\ref{eq13}),} refer to \textcolor{black}{Extended Data Table \ref{table4}}.

In this study, we focus on  \textcolor{black}{the cascade of geostrophic KE}. We defined the small-scale geostrophic KE as
\begin{equation}
	KE_{S}^{g}=\frac{1}{2}\rho _0\tau _{\ell}\left( \mathbf{u^g},\mathbf{u^g} \right),
\end{equation}	
\textcolor{black}{where $\mathbf{u^g}$ denotes geostrophic current velocities. }Consider \textcolor{black}{the following decomposition}: ${KE_{S}} = {KE_{S}^{g}} + {KE_{S}^{ag}}, {J^{flow}} = {J^{g,flow}} + {J^{ag,flow}}, {J^{pre}} = {J^{g,pre}} + {J^{ag,pre}}, {\Pi} = {\Pi^{g}} + {\Pi^{ag}}$ and $ {WP} = {WP^{g}} + {WP^{ag}} $. Here the superscripts “g” and “ag” denote geostrophic and ageostrophic parts, respectively. Through substituting  \textcolor{black}{the above decomposition into Equation (\ref{eq13}), we obtain the budget equation for small-scale geostrophic KE:}

\begin{equation}
	\begin{aligned}
		\frac{{\partial K{E_S}^g}}{{\partial t}} =- J^{g,flow} - J^{g,pre} + \Pi^g  + WP^g + OT.
		\label{eq15}
	\end{aligned}
\end{equation}
\textcolor{black}{The ageostrophic components and the residual terms that are challenging to calculate are all contained in the $OT$ term.} \textcolor{black}{For details of each term in  Equation (\ref{eq15}), refer to  Extended Data Table \ref{table3}.}

\subsection*{\textcolor{black}{Change of KE budget}}
The $Amp$ change can be directly inferred from Equation (\ref{eq15}):
\begin{equation}
	\begin{aligned}
		\Delta Amp =&  - \Delta {\Pi ^g}{|_{{L_{Amp}}}}
		\\
		&= \Delta \left( { - {J^{g,flow}} - {J^{g,pre}} + E{P^g} - \frac{{\partial K{E_S}^g}}{{\partial t}} + OT} \right){|_{{L_{Amp}}}}.
		\label{eq10}
	     \end{aligned}
\end{equation}
Here $\varDelta \mid_{L_{Amp}}^{}$ represents the change in the time-averaged term at the spatial scale $L_{Amp}$ (comparing 2006-2017 to 1994-2005). Refer to \textcolor{black}{Extended Data Table \ref{table3}} for the physical meanings of each term.

\section*{Data Availability}
\textcolor{black}{In this study, the geostrophic current velocity data from satellite altimetry} can be downloaded from Copernicus Marine Environment Monitoring Service at \href{https://doi.org/10.48670/moi-00148}{https://doi.org/10.48670/moi-00148}. CCMP Version-3.1 vector wind analyses are produced by Remote Sensing Systems at  \href{www.remss.com}{www.remss.com}.  The ARMOR3D observation dataset is provided at \href{https://doi.org/10.48670/moi-00052}{https://doi.org/10.48670/moi-00052}. 

\section*{Code availability}
\textcolor{black}{The source codes used for the calculations in this paper are available at \href{https://doi.org/10.5281/zenodo.14581033}{https://doi.org/10.5281/zenodo.14581033} (ref. \cite{jyn_2024_14581033})  and can also be obtained from the corresponding author upon request.}

\section*{Supplementary Information}
\subsection*{Derivation of kinetic energy budget}
Here we show the derivation procedure of the small-scale kinetic energy (KE) budget, which is used  to assess the mechanism of the KE cascade trend. This small-scale KE budget can be obtained through subtracting the large-scale KE budget \cite{Aluie2018Mapping} from the low-pass filtered total KE budget.

\subsubsection*{Derivation of large-scale kinetic energy budget}
\textcolor{black}{The derivation of the large-scale KE budget largely follows ref.  \cite{Aluie2018Mapping}\textcolor{black}{, which starts from the momentum equation,}}
\begin{equation}
\begin{aligned}
	\frac{\partial \mathbf{u}}{\partial t}+\nabla \cdot \left( \mathbf{uu} \right) =-\frac{1}{\rho _0}\nabla P-f\times \mathbf{u}_h+\mathbf{g}+A_h\nabla _{h}^{2}\mathbf{u}+A_z\frac{\partial ^2\mathbf{u}}{\partial z^2}+\mathbf{F}^{forcing},   		
	\label{eqq1}
\end{aligned}
\end{equation}
where $\mathbf{u}$ and $\mathbf{u}_h$ denote 3-dimensional velocity vector and horizontal velocity vector, respectively. $\nabla$ is 3-dimensional divergence operator and $\nabla_h$ is horizontal divergence operator. $P$ denotes pressure, $\rho _0$ is seawater referenced density, $f$ is the Coriolis frequency, $g$ is the gravitational acceleration and $\mathbf{F}^{forcing}$ is the tidal forcing \cite{cushman2011introduction}. $A_h$ and $A_z$ represents the horizontal and vertical viscosity, respectively.

Following ref. \cite{Aluie2018Mapping}, the large-scale KE budget (at scales  $>\ell$) can be obtained from the momentum equation [Equation (\ref{eqq1})] for large-scale flows: 
\begin{equation}
\frac{\partial KE_L}{\partial t}+{J}^{flow}_L=-{J}^{pre}_L-\varPi +\rho _{\ell}\mathbf{g}\cdot \mathbf{u}_{\mathbf{\ell }}+\rho _0A_h\mathbf{u}_{\mathbf{\ell }}\cdot \nabla _{h}^{2}\mathbf{u}_{\mathbf{\ell }}+\rho _0A_z\mathbf{u}_{\mathbf{\ell }}\cdot \frac{\partial ^2\mathbf{u}_{\mathbf{\ell }}}{\partial z^2}+\rho _0\mathbf{F}_{\ell}^{forcing}\cdot \mathbf{u}_{\ell}.
\label{eqq2}
\end{equation}
\textcolor{black}{Here $KE_L$ denotes kinetic energy contained in oceanic motions with spatial  scales larger than $\ell$,}
\begin{equation}
KE_L=\frac{1}{2}\rho _0 \mathbf{u}_{\ell}\cdot  \mathbf{u}_{\ell}.
\end{equation}	
\textcolor{black}{The KE cascade term $\varPi(\ell)$ measures KE transferred from scales $>\ell$ to $<\ell$ due to nonlinear interactions between oceanic motions, and it  can be diagnosed from  \cite{Aluie2018Mapping}:}
\begin{equation}
\varPi(\ell)=-\rho _0\mathbf{S}_{\ell}:\tau _{\ell}\left( \mathbf{u},\mathbf{u} \right) ,
\label{eqpi}
\end{equation}	
where the large-scale strain tensor $\mathbf{S}_{\ell}$ is defined as $\mathbf{S}_{\ell}=\left( \nabla \mathbf{u}_{\ell}+\nabla {\mathbf{u}_{\ell}}^T \right) /2$, and the colon : is a inner product. The subfilter stress tensor,
\begin{equation}
\tau _{\ell}\left( \mathbf{u},\mathbf{u} \right) =\left( \mathbf{uu} \right) _{\ell}-\mathbf{u}_{\ell}\mathbf{u}_{\ell},
\end{equation}	
quantifies the \textcolor{black}{forces exerted by small-scale motions (at scales $<\ell$) on large-scale motions (at scales $>\ell$).}

The term $J_{L}^{flow}$ include two subterms: the former refers to the advection of $KE_L$ by large-scale flows, \textcolor{black}{and the latter denotes the role of small-scale motions in transporting $KE_L$:}
\begin{equation}
J_{L}^{flow}=\nabla \cdot \left( \mathbf{u}_{\ell}KE_L+\rho _0\mathbf{u}_{\ell}^{}\cdot \tau _{\ell}\left( \mathbf{u},\mathbf{u} \right) \right) .
\end{equation}	
The term $J_{L}^{pre}$ denotes the transport of $KE_L$ due to pressure:
\begin{equation}
J_{L}^{pre}=\nabla \cdot \left( \mathbf{u}_{\ell}P_{\ell} \right).
\end{equation}	

Near the ocean surface, the viscosity term  $\rho _0A_h\mathbf{u}_{\mathbf{\ell }}\cdot \nabla _{h}^{2}\mathbf{u}_{\mathbf{\ell }}+\rho _0A_z\mathbf{u}_{\mathbf{\ell }}\cdot \frac{\partial ^2\mathbf{u}_{\mathbf{\ell }}}{\partial z^2}
$ in Equation (\ref{eqq2}) can be divided into two parts:  large-scale wind power input and large-scale oceanic mixing ($Diff_{\ell}$) components \cite{2019Mesoscale}:
\begin{equation}
\rho _0A_h\mathbf{u}_{\mathbf{\ell }}\cdot \nabla _{h}^{2}\mathbf{u}_{\mathbf{\ell }}+\rho _0A_z\mathbf{u}_{\mathbf{\ell }}\cdot \frac{\partial ^2\mathbf{u}_{\mathbf{\ell }}}{\partial z^2}=\underbrace{\left( \rho _0A_h \mathbf{u}_{\mathbf{\ell }}\cdot {\nabla _h}^2\mathbf{u}_{\mathbf{\ell }}+\rho _0A_z \mathbf{u}_{\mathbf{\ell }}\cdot \frac{\partial ^2\mathbf{u}_{\mathbf{\ell }}}{\partial z^2}\mid_{ocean}^{} \right)}_{Diff_{\ell}} +\mathbf{u}_{\mathbf{\ell }}\mid_{surface}^{}\cdot \frac{\tau _{\ell}^{wind}}{H},
\label{eqq6}
\end{equation}	
where \textcolor{black}{$\tau _{\ell}^{wind}$ represents the large-scale wind stress vector (at scales$>\ell$)} and $H$ is approximately the Ekman layer thickness. Therefore, the large-scale KE budget equation can be further written as:
\begin{equation}
\frac{\partial KE_{L}}{\partial t}+{J}^{flow}_{L}=-{J}^{pre}_{L}-\varPi+\rho _{\ell}\mathbf{g}\cdot \mathbf{u}_{\mathbf{\ell }}+Diff_{\ell}+ \frac{\mathbf{u}_{\mathbf{\ell }}\cdot\tau _{\ell}^{wind}}{H}+\rho _0\mathbf{F}_{\ell}^{forcing}\cdot \mathbf{u}_{\ell}.
\label{eqq3}
%	\label{eq8}
\end{equation}

\subsubsection*{Derivation of total kinetic energy budget}
\textcolor{black}{From Equation (\ref{eqq1}), we can derive a governing budget equation for total KE through multiplying Equation (\ref{eqq1}) by $\rho_0\mathbf{u}$:}
\begin{equation}
\begin{aligned}
	\textcolor{black}{	\frac{\partial KE_T}{\partial t}+J_{T}^{flow}=-J_{T}^{pre}+\rho _0\mathbf{g}\cdot \mathbf{u}+Diff_T+\frac{\mathbf{\tau }^{wind}\cdot \mathbf{u}}{H}+\rho _0\mathbf{F}_{}^{forcing}\cdot \mathbf{u}.}
	\label{eqq7}
\end{aligned}
\end{equation}
\textcolor{black}{Here $KE_T$ represents total kinetic energy:}
\begin{equation}
KE_{T}=\frac{1}{2}\rho _0 \mathbf{u}\cdot  \mathbf{u}.
\end{equation}
\textcolor{black}{The term $J_{T}^{flow}$ represents the advection of $KE_T$ by total flow:}
\begin{equation}
J_{T}^{flow}=\nabla \cdot \left( \mathbf{u}KE_T \right), 
\end{equation}
\textcolor{black}{and the term $J_{T}^{pre}$ represents the pressure-induced transport:}
\begin{equation}
\textcolor{black}{J_{T}^{pre}=\nabla \cdot \left( \mathbf{u}P \right).}
\end{equation}
\textcolor{black}{In analogy to  Equation (\ref{eqq6}),  the change of $KE_T$ by viscosity can be divided into $Diff_T$ and total wind power input \cite{2019Mesoscale},} 
\begin{equation}
\rho _0A_h\mathbf{u}\cdot \nabla _{h}^{2}\mathbf{u}+\rho _0A_z\mathbf{u}\cdot \frac{\partial ^2\mathbf{u}}{\partial z^2}=\underset{Diff_T}{\underbrace{\left( \rho _0A_h\mathbf{u}\cdot {\nabla _h}^2\mathbf{u}+\rho _0A_z\mathbf{u}\cdot \frac{\partial ^2\mathbf{u}}{\partial z^2}\mid _{ocean}^{} \right) }}+\frac{\mathbf{u}\mid_{surface}\cdot \mathbf{\tau }^{wind}}{H},
\label{eqq8}
\end{equation}	
where $\mathbf{\tau }^{wind}$ denotes wind stress.

\subsubsection*{Derivation of small-scale kinetic energy budget}
\textcolor{black}{Subtracting the large-scale KE from total KE \textcolor{black}{at a specific filter scale $\ell$}, we obtain the small-scale KE,}
%Considering that the large-scale KE is defined as $KE_{L}=\frac{1}{2}\rho _0 \mathbf{u}_{\ell}\cdot  \mathbf{u}_{\ell}$, the small-scale KE is thus defined as
\begin{equation}
\textcolor{black}{KE_S={(KE_T)}_{\ell}-KE_L=\frac{1}{2}\rho _0\tau _{\ell}\left( \mathbf{u},\mathbf{u} \right),}
\end{equation}
where the subfilter stress $\tau _{\ell}\left( \mathbf{u},\mathbf{u} \right) =\left( \mathbf{uu} \right) _{\ell}-\mathbf{u}_{\ell}\mathbf{u}_{\ell}$\cite{Aluie2018Mapping}. 
\textcolor{black}{The advantage of this $KE_S$ definition is that, at a specific spatial scale, $KE_{L}+KE_{S}$ is equal to total KE and thus no KE leakage issue exists\cite{zhou2021errors,steinberg2022seasonality}. Consistently, subtracting Equation (\ref{eqq3}) from the low-pass filtered version of Equation (\ref{eqq7}) leads to the small-scale KE budget,} 
\begin{equation}
\begin{aligned}
	\frac{{\partial K{E_S}}}{{\partial t}} =  &- \underbrace {\nabla  \cdot \left[ {{{\left( {{\bf{u}}K{E_T}} \right)}_\ell } - {{\bf{u}}_\ell }K{E_L} - {\rho _0}{\bf{u}}_\ell ^{} \cdot {\tau _\ell }\left( {{\bf{u}},{\bf{u}}} \right)} \right]}_{{J^{flow}}} - \underbrace {\nabla  \cdot {\tau _\ell }\left( {{\bf{u}},P} \right)}_{{J^{pre}}} + \Pi  + \underbrace {\frac{{{{\left( {{\bf{\tau }}_{}^{wind} \cdot {\bf{u}}} \right)}_\ell } - {\bf{\tau }}_{\ell}^{wind} \cdot {{\bf{u}}_\ell }}}{H}}_{WP}
	\\
	&+ {\left( {{\rho _0}{\bf{g}} \cdot {\bf{u}}} \right)_\ell } - {\rho _\ell }{\bf{g}} \cdot {{\bf{u}}_\ell } + {\left( {Dif{f_T}} \right)_\ell } - Dif{f_\ell } + {\rho _0}{\tau _\ell }\left( {{\bf{u}},{\bf{F}}_{}^{forcing}} \right).
	\label{eqs1}
\end{aligned}
\end{equation}
\textcolor{black}{In Equation (\ref{eqs1}), the terms $J^{flow}$ and $J^{pre}$ respectively represent the transport of $KE_S$ due to multiscale flows and pressure. The term $\Pi$ represents KE cascade from $KE_L$ to $KE_S$ [Equations  (\ref{eqpi}) and  (\ref{eqq3})].} The term $WP$ represents wind power input into small-scale KE, which can quantify eddy killing. For physical meanings of other terms, refer to Extended Data Table 2 in the main manuscript.

\textcolor{black}{In this study, we focus on geostrophic KE cascade. The  KE contained in small-scale geostrophic flow is},
\begin{equation}
KE_{S}^{g}=\frac{1}{2}\rho _0\tau _{\ell}\left( \mathbf{u^g},\mathbf{u^g} \right),
\end{equation}	
where $\mathbf{u}^{g}$ represents geostrophic velocity. We consider the following geostrophic decomposition: ${KE_{S}} = {KE_{S}^{g}} + {KE_{S}^{ag}}, {J^{flow}} = {J^{g,flow}} + {J^{ag,flow}}, {J^{pre}} = {J^{g,pre}} + {J^{ag,pre}}, {\Pi} = {\Pi^{g}} + {\Pi^{ag}}$ and $ {WP} = {WP^{g}} + {WP^{ag}} $. \textcolor{black}{The superscripts “g” and “ag” in this study denote geostrophic and ageostrophic parts, respectively. Then  we can obtain the budget equation for $KE_{S}^{g}$ by substituting  the above decomposition into Equation (\ref{eqs1}),
\begin{equation}
	\frac{{\partial K{E^g_S}}}{{\partial t}} =- J^{g,flow} - J^{g,pre} + \Pi^g  + WP^g + OT,
	\label{eqq4}
\end{equation}
where the term $OT$ is defined as,
\begin{equation}
	\begin{aligned}
		OT=&- {J^{ag,flow}} - {J^{ag,pre}} + {\Pi ^{ag}} + W{P^{ag}} - \frac{{\partial KE_S^{ag}}}{{\partial t}}\\
		&+ {\left( {{\rho _0}{\bf{g}} \cdot {\bf{u}}} \right)_\ell } - {\rho _\ell }{\bf{g}} \cdot {{\bf{u}}_\ell } + {\left( {Dif{f_T}} \right)_\ell } - Dif{f_\ell } + {\rho _0}{\tau _\ell }\left( {{\bf{u}},{\bf{F}}_{}^{forcing}} \right).
		\label{eqq5}
	\end{aligned}	
\end{equation}}

\textcolor{black}{The small-scale geostrophic KE budget, shown in Equation (\ref{eqq4}), is discussed for each region in the main text.}  The terms on the right side of Equation (\ref{eqq4}) represent the change of $KE_{S}^{g}$ through a range of mechanisms:
The term $J^{g,flow}$ is the geostrophic component of $J^{flow}$, which represents the transport of small-scale KE by geostrophic flow:
\begin{equation}
J^{g,flow}=\nabla_h  \cdot \left[ {{{\left( {{{\bf{u}}^g}KE_T^g} \right)}_\ell } - {\bf{u}}_\ell ^gKE_L^g - {\rho _0}{\bf{u}}_\ell ^g \cdot {\tau _\ell }\left( {{{\bf{u}}^g},{{\bf{u}}^g}} \right)} \right],
\end{equation}	
where $\nabla _h$ is the horizontal divergence operator. The term $J^{g,pre}$ denotes pressure work on small-scale KE by geostrophic flow:
\begin{equation}
J^{g,pre}=\nabla_h  \cdot {\tau _\ell }\left( {{{\bf{u}}^g},P} \right).
\end{equation}	
The term $\varPi^{g}$ \textcolor{black}{denotes the contribution of geostrophic motions to KE cascade}:
\begin{equation}
\varPi^{g} =-\rho _0\left[ \tau _{\ell}\left( u^g,u^g \right) \frac{\partial u_{\ell}^{g}}{\partial x} + \tau _{\ell}\left( u^g,v^g \right) \left( \frac{\partial u_{\ell}^{g}}{\partial y} + \frac{\partial v_{\ell}^{g}}{\partial x} \right) + \tau _{\ell}\left( v^g,v^g \right) \frac{\partial v_{\ell}^{g}}{\partial y} \right], 
%	\label{eq1}
\end{equation}
where $u$ and $v$ denote the zonal and meridional current velocities, respectively.
The term $WP^g$ represents wind power input into small-scale KE due to geostrophic motions:
\begin{equation}
WP^{g}=\frac{{{{\left( {{\bf{\tau }}_{}^{wind} \cdot {{\bf{u}}^g}} \right)}_\ell } - {\bf{\tau }}_{\ell}^{wind} \cdot {\bf{u}}_\ell ^g}}{H}.
\end{equation}

\textcolor{black}{The last term $OT$, i.e., $- {J^{ag,flow}} - {J^{ag,pre}} + {\Pi ^{ag}} + W{P^{ag}} - \frac{{\partial KE_S^{ag}}}{{\partial t}}+ {\left( {{\rho _0}{\bf{g}} \cdot {\bf{u}}} \right)_\ell } - {\rho _\ell }{\bf{g}} \cdot {{\bf{u}}_\ell } + {\left( {Dif{f_T}} \right)_\ell } - Dif{f_\ell } + {\rho _0}{\tau _\ell }\left( {{\bf{u}},{\bf{F}}_{}^{forcing}} \right)$, includes the components challenging to calculate directly from observations.  Therefore, this study chose to estimate $OT$ in each region by calculating all the other terms in Equation (\ref{eqq4}) first and then closing the budget equation. As shown in Extended Data Table 3 in the main manuscript, the term ${J}^{ag,flow}$ represents transport of small-scale KE by ageostrophic flow, while ${J}^{ag,pre}$ represents pressure work on $KE_S$ by ageostrophic flow. ${\Pi}^{ag}$ is the contribution of ageostrophic motions to KE cascade, and $WP^{ag}$ represents wind power input into $KE_S$ due to ageostrophic motions. $\frac{{\partial KE_S^{ag}}}{{\partial t}}$ is the tendency term of small-scale ageostrophic KE. Other terms in $OT$ include the energy transfer between small-scale available potential energy and small-scale KE (${\left( {{\rho _0}{\bf{g}} \cdot {\bf{u}}} \right)_\ell } - {\rho _\ell }{\bf{g}} \cdot {{\bf{u}}_\ell }$), diffusion of small-scale KE (${\left( {Dif{f_T}} \right)_\ell } - Dif{f_\ell }$), and energy transfer to $KE_S$ by tidal forcing (${\rho _0}{\tau _\ell }\left( {{\bf{u}},{\bf{F}}_{}^{forcing}}\right)$).  }

\newpage
%-----------------------------------enter bibliography here--------------------------------------------------------------------------------------------------------- 
%\bibliography{reference1}
\bibliographystyle{plain}

\section*{Acknowledgements}
\textcolor{black}{This study was supported by National Natural Science Foundation of China (42476007, 42076007).}

\section*{Author contributions}
R.C. conceived and oversaw this study,  Q. G. conducted the analysis and wrote the initial manuscript,  B. Q. and Z. J. contributed by offering alternative interpretations or analysis suggestions. X. S. and Y. C. contributed about the diagnosis techniques. All the authors reviewed and edited the manuscript.

\section*{Competing interests}
The authors declare no competing interests.

\appendix
\newpage
\renewcommand\thetable{\arabic{table}}  
\renewcommand{\tablename}{Extended Data Table}

\setcounter{table}{0}
\counterwithin*{table}{part}
\begin{table}[!htbp]
	\caption{\textcolor{black}{\textbf{QG model parameters.}} "Lat" denotes the latitudes for each region. \textcolor{black}{All the model runs reached equilibrium after the spin up time.}}\label{table2}
	\centering
	\begin{adjustbox}{width=\textwidth}
		\begin{tabular}{|c|c|c|c|c|c|c|c|c|}
			\hline
			\textbf{Region} & 
			\textbf{Model runs} & 
			$\delta$ & 
			$U_1$ \textbf{(m/s)} &
			$U_2$ \textbf{(m/s)} & 
			$U_1 - U_2$ \textbf{(m/s)} & 
			$R_d$ \textbf{(km)} & 
			\textbf{Lat (°N)} & 
			\textbf{Spin up time (years)} \\ \hline
			\multirow{5}{*}{\textbf{KAE}} & I & 0.1700 & 0.1300 & 0.0700 & 0.0600 & 36.00 & 40 & 200 \\ \cline{2-9} 
			& II & 0.1750 & 0.1210 & 0.0622 & 0.0588 & 36.07 & 40 & 200 \\ \cline{2-9} 
			& III & 0.1750 & 0.1300 & 0.0700 & 0.0600 & 36.00 & 40 & 200 \\ \cline{2-9} 
			& IV & 0.1700 & 0.1210 & 0.0622 & 0.0588 & 36.00 & 40 & 200 \\ \cline{2-9} 
			& V & 0.1700 & 0.1300 & 0.0700 & 0.0600 & 36.07 & 40 & 200 \\ \hline
			\multirow{5}{*}{\textbf{GS}} & I & 0.1700 & 0.1000 & 0.0600 & 0.0400 & 30.00 & 40 & 500 \\ \cline{2-9} 
			& II & 0.1828 & 0.0941 & 0.0545 & 0.0396 & 30.06 & 40 & 500 \\ \cline{2-9} 
			& III & 0.1828 & 0.1000 & 0.0600 & 0.0400 & 30.00 & 40 & 500 \\ \cline{2-9} 
			& IV & 0.1700 & 0.0941 & 0.0545 & 0.0396 & 30.00 & 40 & 500 \\ \cline{2-9} 
			& V & 0.1700 & 0.1000 & 0.0600 & 0.0400 & 30.06 & 40 & 500 \\ \hline
			\multirow{5}{*}{\textbf{ACC}} & I & 0.2100 & 0.1600 & 0.1300 & 0.0300 & 20.00 & -45 & 200 \\ \cline{2-9} 
			& II & 0.2174 & 0.1621 & 0.1313 & 0.0308 & 20.10 & -45 & 200 \\ \cline{2-9} 
			& III & 0.2174 & 0.1600 & 0.1300 & 0.0300 & 20.00 & -45 & 200 \\ \cline{2-9} 
			& IV & 0.2100 & 0.1621 & 0.1313 & 0.0308 & 20.00 & -45 & 200 \\ \cline{2-9} 
			& V & 0.2100 & 0.1600 & 0.1300 & 0.0300 & 20.10 & -45 & 200 \\ \hline
		\end{tabular}
	\end{adjustbox}
\end{table}

\newpage
\begin{table}[!htbp]
	\caption{\textbf{\textcolor{black}{Interpretation of the small-scale KE budget [Equation (\ref{eq13})].}} $\mathbf{u}$ denotes the 3-dimensional current velocity vector. $\rho _0$ is the seawater referenced density. $\nabla _h$ and $\nabla$ represent the horizontal and 3-dimensional divergence operators, respectively. \textcolor{black}{$A_h$ and $A_z$ are horizontal and vertical viscosity, respectively.} \textcolor{black}{The colon (:) denotes an inner product.} }\label{table4}
	\centering
	\begin{adjustbox}{width=0.7\textwidth}	
		\begin{tabular}{|c|c|c|}
			\hline
			\textbf{Term} & \textbf{Mathematical form} & \textbf{Meaning} \\ \hline
			$KE_T$ & $\frac{1}{2}\rho _0\mathbf{u}^2$ & Total KE \\ \hline		
			$KE_{L}$ & $\frac{1}{2}\rho _0{\mathbf{u}_{\ell}}^2$ & Large-scale KE \\ \hline	
			$KE_{S}$ & $(KE_T)_{\ell}-KE_L$ & Small-scale KE \\ \hline							
			$\left( Diff_T \right) _{\ell}$ & $\left(  \rho _0A_h\mathbf{u}\cdot {\nabla^2 _h}\mathbf{u}+\rho _0A_z\mathbf{u}\cdot \frac{\partial ^2\mathbf{u}}{\partial z^2}\mid _{ocean}^{} \right) _{\ell}$ & Large-scale component of $KE_T$ diffusion \\ \hline
			$Diff_{\ell}$ & $\rho _0A_h \mathbf{u}_{\mathbf{\ell }}\cdot {\nabla^2 _h}\mathbf{u}_{\mathbf{\ell }}+\rho _0A_z \mathbf{u}_{\mathbf{\ell }}\cdot \frac{\partial ^2\mathbf{u}_{\mathbf{\ell }}}{\partial z^2}\mid_{ocean}^{}$ & Diffusion of $KE_L$ \\ \hline
			$\varPi$ & $-\rho _0\mathbf{S}_{\ell}:\tau _{\ell}\left( \mathbf{u},\mathbf{u} \right)$, $\mathbf{S}_{\ell}=\left( \nabla \mathbf{u}_{\ell}+\nabla {\mathbf{u}_{\ell}}^T \right) /2$ & KE cascade from $KE_L$ to $KE_S$\\ \hline
			
		\end{tabular}
	\end{adjustbox}
\end{table}

\newpage
\begin{table}[!htbp]
	\caption{\textbf{\textcolor{black}{Interpretation of the small-scale geostrophic KE budget [Equation (\ref{eq15})]}.} $\nabla _h$ is the horizontal divergence operator. $u$ and $v$ are zonal and meridional velocities, respectively. The superscript “g” represents geostrophic components. $EAPE$ represents available eddy potential energy. $EKE$ represents eddy kinetic energy.} \label{table3}
	\centering
	\begin{adjustbox}{width=\textwidth}	
		\begin{tabular}{|c|c|c|}
			\hline
			\textbf{Term} & \textbf{Mathematical form} & \textbf{Meaning} \\ \hline
				$KE_{S}^{g}$ & $\frac{1}{2}\rho _0\left[ (\mathbf{u}^g\mathbf{u}^g)_{\ell}-\mathbf{u}_{\ell}^g \mathbf{u}_{\ell}^g \right]$ & KE contained in small-scale geostrophic motions \\ \hline		
			$\varPi ^{g}$ & $-\rho _0\left[ \tau _{\ell}\left( u^g,u^g \right) \frac{\partial u_{\ell}^{g}}{\partial x} + \tau _{\ell}\left( u^g,v^g \right) \left( \frac{\partial u_{\ell}^{g}}{\partial y} + \frac{\partial v_{\ell}^{g}}{\partial x} \right) + \tau _{\ell}\left( v^g,v^g \right) \frac{\partial v_{\ell}^{g}}{\partial y} \right]$ & Contribution of geostrophic motions to KE cascade \\ \hline				
			$J^{g,flow}$ & $\nabla_h  \cdot \left[ {{{\left( {{{\bf{u}}^g}KE_T^g} \right)}_\ell } - {\bf{u}}_\ell ^gKE_L^g - {\rho _0}{\bf{u}}_\ell ^g \cdot {\tau _\ell }\left( {{{\bf{u}}^g},{{\bf{u}}^g}} \right)} \right]$ & \makecell*[c]{Transport of small-scale KE by geostrophic flow} \\ \hline
			$J^{g,pre}$ & $\nabla_h  \cdot {\tau _\ell }\left( {{{\bf{u}}^g},P} \right)$ & Pressure work on small-scale KE by geostrophic flow \\ \hline
			$WP^{g}$ & \makecell*[c]{$\frac{{{{\left( {{\bf{\tau }}_{}^{wind} \cdot {{\bf{u}}^g}} \right)}_\ell } - {\bf{\tau }}_{\ell}^{wind} \cdot {\bf{u}}_\ell ^g}}{H}$} & \makecell*[c]{Wind power input into small-scale KE \\due to geostrophic motions} \\ \hline			
			$OT$ & \makecell*[c]{$\begin{array}{l}
				- {J^{ag,flow}} - {J^{ag,pre}} + {\Pi ^{ag}} + E{P^{ag}} - \frac{{\partial KE_S^{ag}}}{{\partial t}}\\
				+ {\left( {{\rho _0}{\bf{g}} \cdot {\bf{u}}} \right)_\ell } - {\rho _\ell }{\bf{g}} \cdot {{\bf{u}}_\ell } + {\left( {Dif{f_T}} \right)_\ell } - Dif{f_\ell } + {\rho _0}{\tau _\ell }\left( {{\bf{u}},{\bf{F}}_{}^{forcing}} \right)
			\end{array}$}& See below \\	\hline	
			\multirow{9}{*}{\makecell*[c]{Terms \\in $OT$}} 
			& ${J}^{ag,flow}=J^{flow}-J^{g,flow}$ & Transport of small-scale KE by ageostrophic flow\\ \cline{2-3}  
			& ${J}^{ag,pre}=J^{pre}-J^{g,pre}$ & Pressure work on small-scale KE by ageostrophic flow\\ \cline{2-3} 
			& ${\Pi}^{ag}=\Pi-\Pi^{g}$ & Contribution of ageostrophic motions to KE cascade\\ \cline{2-3} 			
			& $WP^{ag}=WP-WP^g$ & \makecell*[c]{Wind power input into small-scale KE \\due to ageostrophic motions} \\ \cline{2-3}
			& \makecell*[c]{$\frac{{\partial KE_S^{ag}}}{{\partial t}}=\frac{{\partial KE_S}}{{\partial t}}-\frac{{\partial KE_S^{g}}}{{\partial t}}$} & \makecell*[c]{Tendency term of  small-scale ageostrophic KE} \\ \cline{2-3} 
			& ${\left( {{\rho _0}{\bf{g}} \cdot {\bf{u}}} \right)_\ell } - {\rho _\ell }{\bf{g}} \cdot {{\bf{u}}_\ell }$ & Energy transfer from small-scale EAPE to small-scale KE\\ \cline{2-3}
			& ${\left( {Dif{f_T}} \right)_\ell } - Dif{f_\ell }$ & Diffusion of small-scale KE \\ \cline{2-3}					
			& ${\rho _0}{\tau _\ell }\left( {{\bf{u}},{\bf{F}}_{}^{forcing}}\right)$ & Energy transfer to small-scale KE by tidal forcing \\ \hline
		\end{tabular}
	\end{adjustbox}
\end{table}

\newpage
\renewcommand\thefigure{\arabic{figure}}  
\renewcommand{\figurename}{Extended Data Fig.}
\setcounter{figure}{0}  
\counterwithin*{figure}{part}
\begin{figure}[ht]
	\centering
	\includegraphics[width=0.9\linewidth]{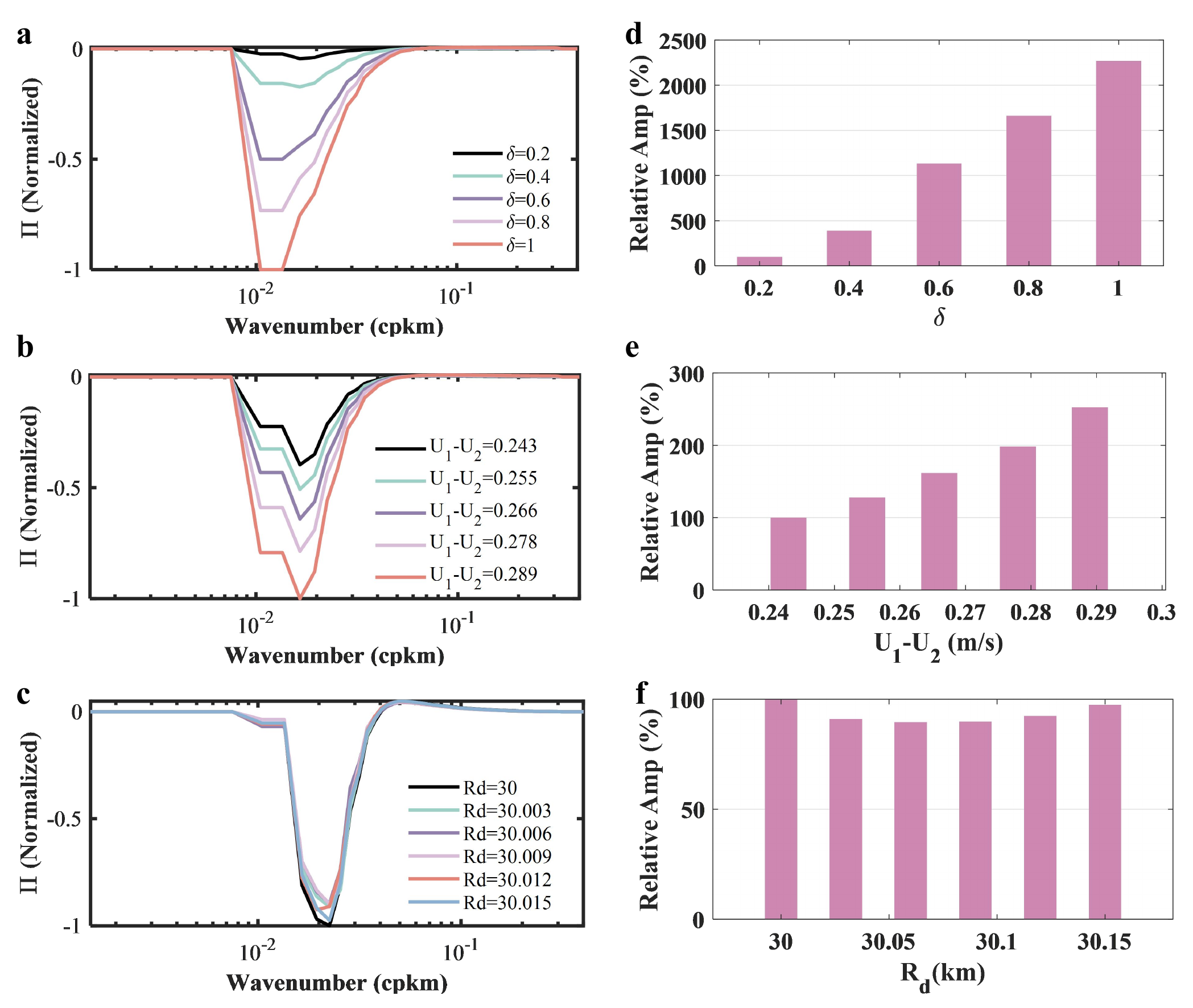}
	\caption{\textcolor{black}{\textbf{Inverse KE cascade response to idealized oceanic changes.} \textbf{a}-\textbf{c}, Upper layer spectral KE fluxes for each QG experiment (normalized by the minimum value from the experiment group in each panel). The experimental scenarios are (a) adjustment of $\delta$, (b) adjustment of $U_1-U_2$, (c) adjustment of $R_d$. \textbf{d}-\textbf{f}, Relative $Amp$ from the experiments in (a) to (c), which is defined as the ratio between $Amp$ in each experiment and that in the baseline experiment.  The baseline conditions are: (d) $\delta=0.2$, $U_1-U_2=0.231 m/s$, $R_d=40 km$; (e) $\delta=0.2$, $U_1-U_2=0.243 m/s$, $R_d=40 km$; (f) $\delta=0.2$, $U_1-U_2=0.058 m/s$, $R_d=30 km$. }}
	\label{extendfig1}
\end{figure}

\newpage
\begin{figure}[!htbp]
	\centering
	\includegraphics[width=0.9\linewidth]{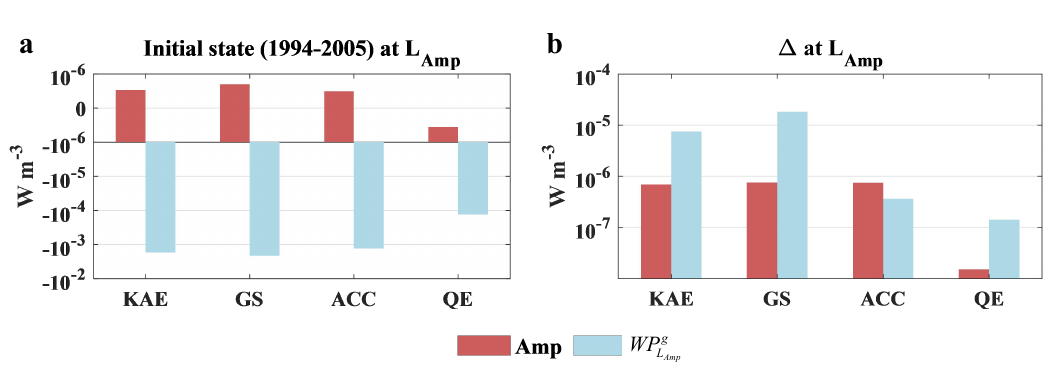}
	\caption{\textcolor{black}{\textbf{Relation between changes in eddy killing and the inverse KE cascade amplitude ($Amp$).} \textbf{a}, Climatological $Amp$ and the wind power input term ($WP^g$ in Equation (\ref{eq15})) at the spatial scale $L_{Amp}$  for the period 1994-2005. \textbf{b}, Differences in $Amp$ and $WP_{L_{Amp}}^g$  between 2006-2017 and 1994-2005. Red (blue) bars represent $Amp$ ($WP_{L_{Amp}}^g$). The Ekman layer depth in the calculation of $WP^g$ is set to 10 meters. The magnitude of $\Delta WP^g$ remains large when the Ekman layer depth ranges from 10 to 50 meters.}}
	\label{extendfig2}
\end{figure}

\end{document}